\documentclass[preprint]{aastex63}
\usepackage{shortvrb}
\usepackage{graphicx}                   
\usepackage{amssymb}                    
\usepackage{color}                       
\usepackage{url}  
\usepackage{epsfig} 

\begin{document}

\title{Temporal and latitudinal variations of Ca-K plage and network area: An implication to meridional flows}

\author{Muthu Priyal} 
\affiliation{Indian Institute of Astrophysics, Bangalore, India}
\author{Jagdev Singh}
\affiliation{Indian Institute of Astrophysics, Bangalore, India}
\author{Belur Ravindra} 
\affiliation{Indian Institute of Astrophysics, Bangalore, India}
\author{G. Sindhuja}
\affiliation{Indian Institute of Astrophysics, Bangalore, India}
\correspondingauthor{Belur Ravindra}
\email{ravindra@iiap.res.in}

\begin{abstract}

The Ca-K spectroheliograms obtained at the Kodaikanal observatory (KO) are used to generate a uniform time series using the equal contrast technique (ECT) to study the long and short-term variation in the solar chromosphere. The percentage of plage, Enhanced network (EN), Active network (AN), and Quiet network (QN) area at various latitudes is compared with the activity at 35$^{\circ}$ latitude and also with the sunspot number for the period of 1907 -- 1984. The values of phase differences indicate that the activity begins at $\sim$45$^{\circ}$ latitude and shift progressively to the lower latitude at a speed of 
$\sim$~9.4~m~sec$^{-1}$ . The shift speed slows down gradually and reaches 
$\sim$~3~m~sec$^{-1}$ at $\sim$5$^{\circ}$  latitude. No phase difference between the variations of Ca-K activity at 55$^{\circ}$, 65$^{\circ}$, and 75$^{\circ}$ latitude belts implies that changes in the activity are happening simultaneously. The analysis shows that the activity at polar latitude belts is anti-correlated with the sunspot number. This study indicates that a multi-cell meridional flow pattern could exist in the solar convection zone. One type of cell could transport the magnetic elements from mid-latitude to low-latitude belts through meridional flows, and the other one could be operating in the polar region. 

\end{abstract}


\section{Introduction}
Almost all the variations on the Sun are related to the growth and decay of the active regions on the Sun. The existence of the solar cycle was discovered by \citet{schwabe1843} from the variations in the sunspot number over 11-years. The sunspots appear at mid-latitudes during the beginning of the cycle and then appear at lower and lower latitudes with time. 

\citet{hale1908} found that sunspots are regions of a high magnetic field. \citet{babcock1961} proposed that differential rotation of the Sun generates toroidal field and decay of sunspots results in the generation of poloidal field that shifts towards the northern and southern poles of the Sun. The meridional flow governs the pattern of occurrence of sunspots on the Sun. \citet{howard1981} suggested that polar field formation is due to the migration of weak magnetic fields toward the poles. The poleward migration of weak magnetic field was also reported by \citet{wang1988, cameron1998} and \citet{durrant2004}. \citet{ananthakrishnan1952} reported the movement of prominences from mid-latitudes towards the poles. \citet{hathaway2003} discussed the role of the equatoward meridional flow in transporting the toroidal component of the magnetic field. Modeling the solar dynamo \citep{dikpati2006, dikpati2010} has shown the importance of the meridional flow pattern for understanding the solar cycle. 

\par The systematic large-scale plasma flows from the equator toward the poles just below the solar surface and from the poles toward the equator deep inside the Sun plays a vital role in the Sun's magnetic dynamo \citep{choudhuri1995, charbonneau2007, nandy2011}. The meridional flow studies made using the magnetic field observations indicated that during the minima phase of the solar cycle, the magnetic field peaked at the polar region \citep{wang1988}. \citet{worden2000} pointed out that measurements of flux distribution beyond the 75$^{\circ}$ latitudes have significant uncertainty because of canopy effects and geometrical foreshortening of features. \citet{raouafi2007} found that the field strength remains flat between the latitude of 55$^{\circ}$ and 75$^{\circ}$ and then decreases by more than 50\% in the polar region. Using the feature tracking technique, the meridional flow speed on the solar surface has been found to be between 10 -- 20~m~sec$^{-1}$ towards the pole \citep{komm1993, hathaway2010}. \citet{zhao2013} reported poleward flow with a velocity of 15~m~sec$^{-1}$ and equator-ward flow in the middle of the convection zone, suggesting a double-cell meridional flow profile. In contrast, a detailed data analysis by \citet{rajaguru2015} suggests a single-cell meridional circulation. 

\citet{leighton1959} reported that the Ca-K plage regions have a magnetic field between 100 to 200 Gauss and Ca-K intensity related to the strength of the magnetic field. \citet{leighton1964} showed that the pattern of Ca-K network cells resembles the boundaries of large convective cells seen in the photospheric dopplergrams. Further, \citet{sivaraman1982} found that Ca-K bright points and weak magnetic field regions are co-spatial and bright region changes their location with the movement of an associated magnetic field. Thus, the variations in Ca-K features or flux can be used to understand the solar dynamo because of the correlation between the Ca-K line and the magnetic field regions at all length and strength scales on the Sun \citep{skumanich1975, ortiz2005}.

\citet{sindhuja2014} analyzed the spectra obtained at Kodaikanal Observatory (KO) as a function of latitude, integrated over the visible longitudes for 1989 – 2011. The details of observations and the initial analysis are given in \citet{singh1989, singh2004}. The increase in K$_{1}$ and decrease in $K_{2}$ width indicates an increase in Ca-K emission, implying an increase in activity \citep{white1982, sivaraman1987}. \citet{sindhuja2014, sindhuja2015} found that K$_{1}$ and K$_{2}$ widths of the Ca-K line varied at all latitudes with 11--year period, but a maximum of K$_{1}$ width occurred at different times at different latitude belts. The K$_{1}$ width varied by ${\sim}$ 30\% at the equatorial belts but only ${\sim}$ 6\% at latitude belts around 50$^{\circ}$. The variations in K$_{1}$ width in the polar region are anti-correlated with the variation at the 35$^{\circ}$ latitude belt by a phase difference of about 5 years, in agreement with the findings of \citet{makarov2004}. From the phase difference at different latitude belts with respect to the 35$^{\circ}$ belt, they found that during the beginning of the solar cycle, the activity shifted from mid-latitude towards the equator with a speed of ${\sim}$ 5~m~s$^{-1}$ in the northern and ${\sim}$ 14~m~s$^{-1}$ in the southern hemisphere. 

Following these investigations, \citet{pooja2021} selected ${\sim}$ 22960 out of 34453 corrected images to identify Ca-K line features such as plage, EN, AN, and QN areas integrated over 10$^{\circ}$ latitude belts by using the intensity and area threshold values. Due to significant data gaps, the resulting parameters were averaged over a month with a running average over three years. Considering the data of 9 solar cycles, they found that activity shifts from mid-latitude towards the equator at a faster speed at the beginning of the solar cycle. The speed decreases as the cycle progresses. The average speed of activity shift varied between ${\sim}$ 19 and 3~m~s$^{-1}$ from 1913 -- 2004.   

\citet{singh2021} have rescaled the Ca-K spectroheliograms (KO) using the Equal Contrast Technique (ECT) to generate a more uniform time series consisting of all the 34453 images taken over 16981 days during the period 1905 -- 2007. There are fewer gaps in the data as compared to the time series used by \citet{pooja2021}. We have analyzed this time series as a function of latitude to study the variation of plage, EN, AN, and QN areas with time. In this article, we report the variations of Ca-K parameters with better accuracy with averaged parameters over a smaller period. We computed the speed of activity shift from mid-latitude towards equator and poles for each solar cycle from 1905 -- 2004.

\section{Data Analysis}

\begin{figure}[!ht]
\centering
\includegraphics[width=0.8\textwidth]{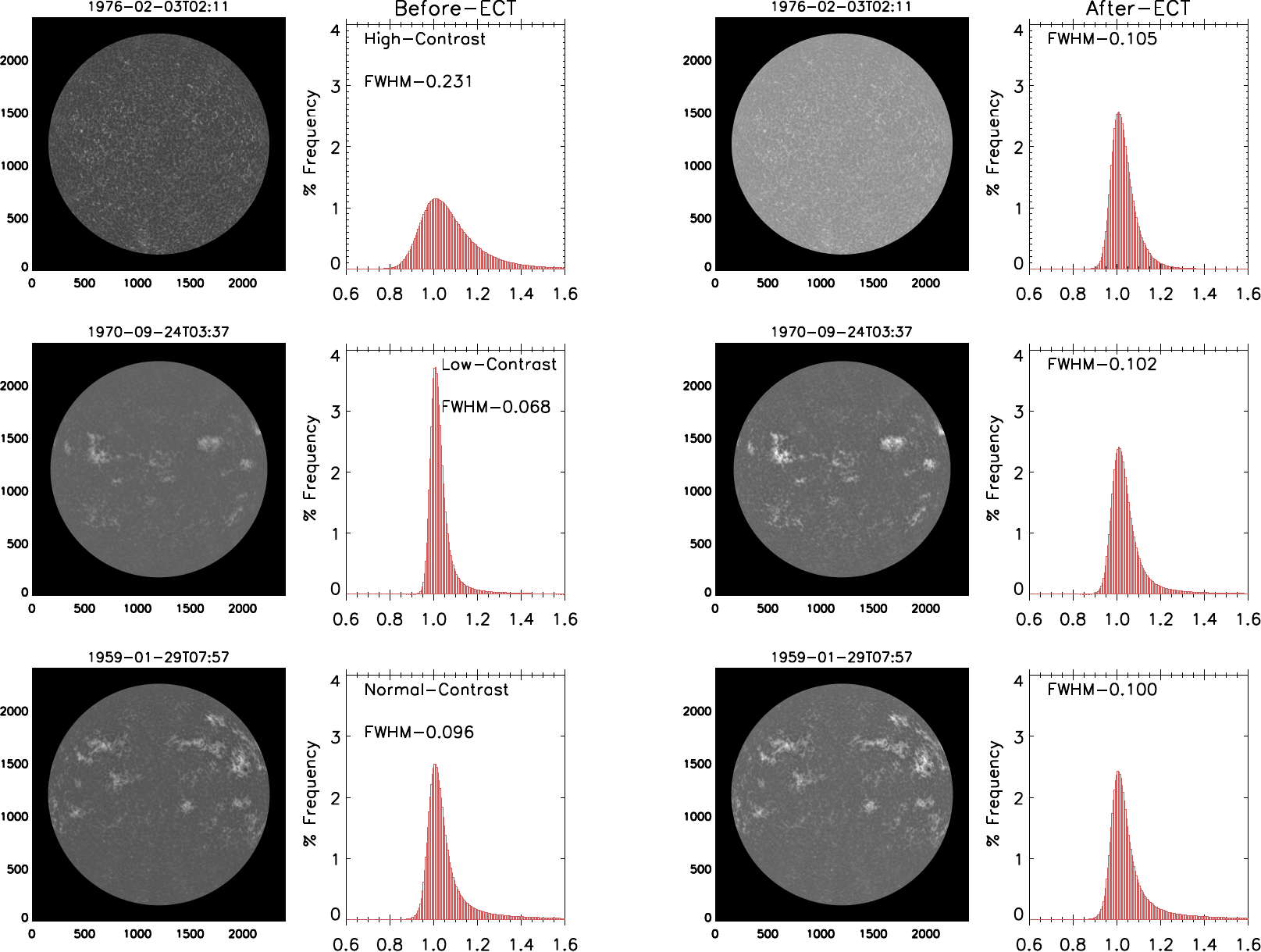}
\caption{Four panels in the first row of the figure show the image before the ECT application, its intensity distribution, image after the ECT procedure and its intensity distribution for a high contrast image obtained during quiet phase of the sun. The FWHM of the intensity distribution is indicated in each panel of intensity distribution. The middle and bottom row show these for the low and normal contrast images, respectively.}
\label{fig:1}
\end{figure}

The Ca-K spectroheliograms were obtained daily at the Kodaikanal observatory from 1905 to 2007, with a few gaps due to sky conditions and instrument problems. These images were digitized with a pixel resolution of 0.86~arcsec and 16-bit readout using a CCD detector of 4K $\times$ 4K format. The digitized images were calibrated, and a long time series of equal contrast images were generated by making full width at half maximum (FWHM) of normalized intensity distribution lying between 0.10 -- 0.11 \citep{priyal2019, singh2021}.  The first and second columns of Figure~\ref{fig:1} shows the images with high (quiet phase, FWHM = 0.231), low (active phase, FWHM = 0.068) and intermediate contrast (FWHM = 0.096) and normalized intensity distribution of these images, respectively. In histograms, the area under the curve with intensity $>$ 1.1 is an indicator of Ca-K active regions. This area for the quiet phase is more than that for the active phase, contrary to the expectation. This is because of the difference in the contrast of these images. The third and fourth columns of the figure show the images and their intensity distribution after applying the ECT procedure. The values of FWHM of the intensity distribution become similar after the ECT application for all the images and hence their contrast.
 
\begin{figure}[!ht]
\centering
\includegraphics[width=0.5\textwidth]{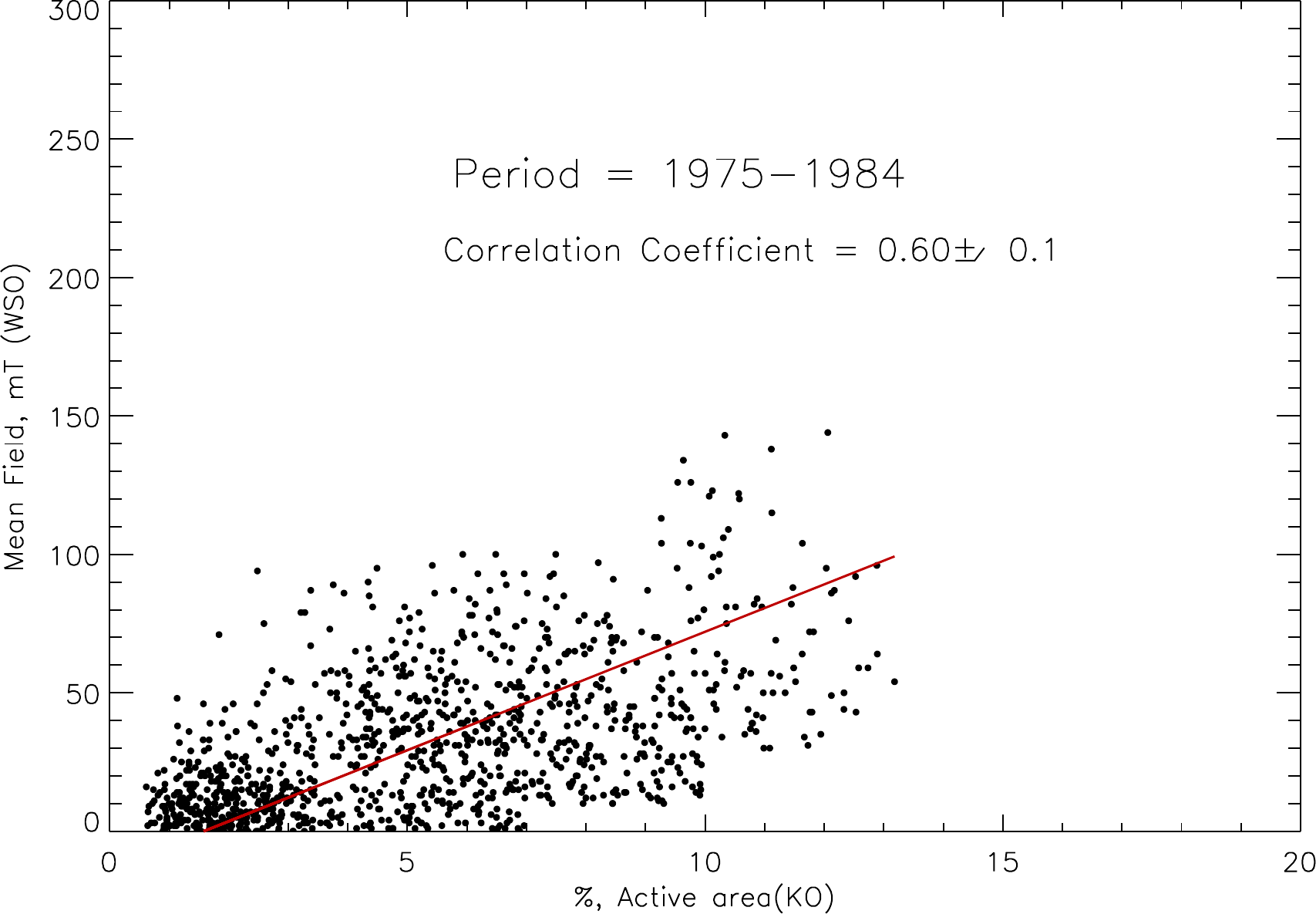}
\caption{Scatter plot of mean magnetic field measured at Wilcox Solar Observatory (WSO) versus the percentage Ca-K active area.}
\label{fig:2}
\end{figure}

\begin{figure}[!ht]
\centering
\includegraphics[width=0.8\textwidth]{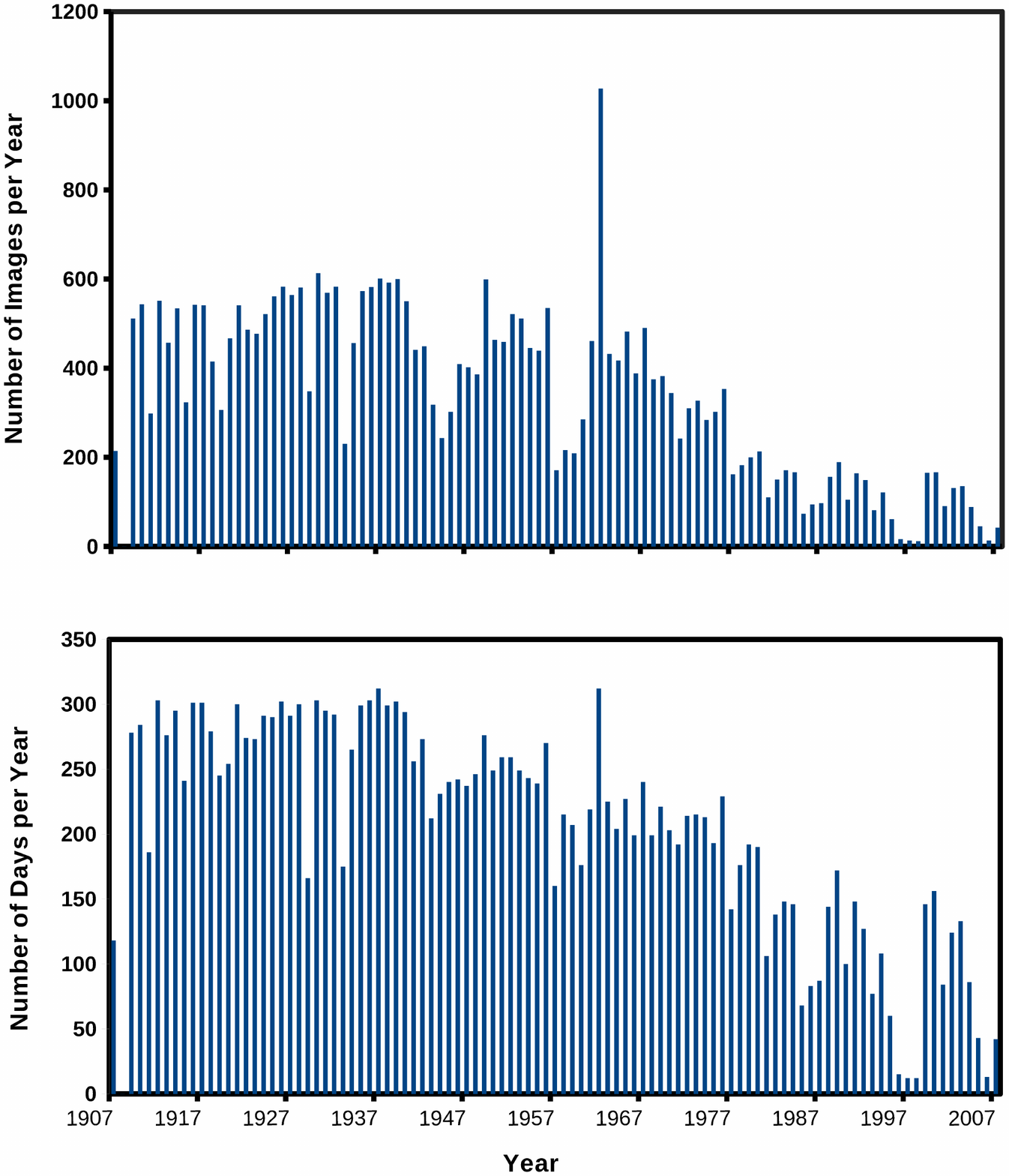}
\caption{Upper panel of the figure shows the number of images per year obtained at the Kodaikanal observatory (KO) during the period 1907 -- 2007. Lower panel indicates the observations made on number of days per year.}
\label{fig:3}
\end{figure}
 
We segregated the Ca-K features such as plages, EN, AN, and QN regions based on the intensity and area threshold. Then, we defined the plage area as regions on the Sun having normalized intensity $>$ 1.30 and an area of consecutive pixels $>$ 500 and EN with area $>$ 10 and $<$500 pixels. The 500 pixels are equivalent to 370 arcsec$^{2}$, about half the size of a large convective cell, and 10 pixels correspond to 7.4~arcsec$^{2}$. The threshold values for AN have been defined as regions with intensity $>$ 1.20 and $<$ 1.30 with size $>$10 pixels and QN with intensity $>$ 1.10 and $<$ 1.20 with an area of 5 pixels equivalent to 3.7~arcsec$^{2}$, little bigger than the size of a granule. Generally, the width of the network boundary is about 8~arcsec equal to 10 pixels \citep{singh1981}. Therefore, a minimum number of pixels to identify EN, AN, and QN was decided to avoid the contribution due to groups of a couple of pixels, those may not be because of Ca-K active region. Considering the radius of the observed image and the B-angle of the Sun on the day of observations, we prepared the solar grid with an interval of 10$^{\circ}$ in latitude up to 90$^{\circ}$. The areas of different Ca-K features are determined in each segment using the intensity and area threshold values. After analyzing all images, the average values of Ca-K parameters are derived in case of multiple images available on single day. We have used the Ca-K data for period 1907 -- 84 to study the phase differences in the variation of active areas at various latitudes. 
 
To see how well the solar mean magnetic field and overall Ca-K brightenings are correlated, we used the magnetic field records of the Wilcox observatory ($http://sun.stanford.edu/wso/meanfld/$). We plot the measured values of Ca-K active area (integrated area over all latitudes) versus magnetic field (Figure~\ref{fig:2}) for the period 1975 -- 84, which shows a good agreement between the two parameters with a correlation coefficient of 0.60.

 Like any other ground-based observatory, KO also receives Monsoon rains between June - October. During this period number of days with a clear sky are less. The upper panel of Figure~\ref{fig:3} shows the histograms of the number of images per year available in the Ca-K time series. The bottom panel of the figure indicates the number of days per year when observations were made. Generally, observations could be made on $>$ 250 days per year until 1957 and $>$ 200 days per year from 1958 to 1983, restricted mainly by the sky conditions. The number of images reduced to $<$ 150 per year from 1984 to 2007. This is because of the weather, the working condition of the instrument, and the availability of photographic films.  The observations could not be obtained on some of the days cause gaps in the data. The derived values of plage, EN, AN, and QN areas were interpolated using the IDL subroutine ``Interpol.pro'' for the non-availability of data on these days by considering the values adjacent to the gap.
 
 \par In this paper, we investigate the variations in area of plages and small scale structures as a function of latitude and time. There are numerous small-scale features found at each latitude belt. We, therefore, have not applied the correction for the foreshortening effect to the area of the detected features. We have just computed the total area of plage and small scale features at each latitude belt of 10$^{\circ}$ each.  To account for the different area of each latitude belt, we have determined the fractional area of each feature by considering the total area of that latitude belt. Then, we determined the time difference between the occurrences of maximum of each feature in the time series, at different latitude belts with respect to 35$^{\circ}$ latitude belt by employing the cross-correlation between the two. Similarly, we computed the time difference between the peaks of Ca-K features in the time series of data at different latitude belts with respect to sunspot number whole disk as a reference. Considering the radius and circumference of the sun at zero degrees longitude, the surface distance between centers of any two latitudes separated by 10$^{\circ}$ is 121475 km, irrespective of the latitude. Thus, phase difference of one month between two latitudes yields a velocity of 47 m/sec on the solar surface. The velocity is computed using the distance in meters on the solar surface divided by the phase difference in seconds, thus, making it independent of the spherical geometry of the sun. We are computing the speed of shift in the latitude direction. Thus, the movement along the longitudinal direction due to spherical geometry may not have any effect on the results.

\section{Results}

We study the variation in detected fractional plage area and small scale feature (EN, AN, and QN) with time at various latitudes at an interval of 10$^{\circ}$ up to 80$^{\circ}$ latitude belts. Sunspots appear at ${\sim}$ 35$^{\circ}$ latitudes at the beginning of the solar cycle.  Therefore, we determine the speed of shift of activity by computing the phase difference between the occurrences of maximum activity at various latitudes with respect to that at 35$^{\circ}$ latitude belt.

\subsection{Latitudinal variation of fractional plage area with time}

\begin{figure}[!ht]
\centering
\includegraphics[width=0.8\textwidth]{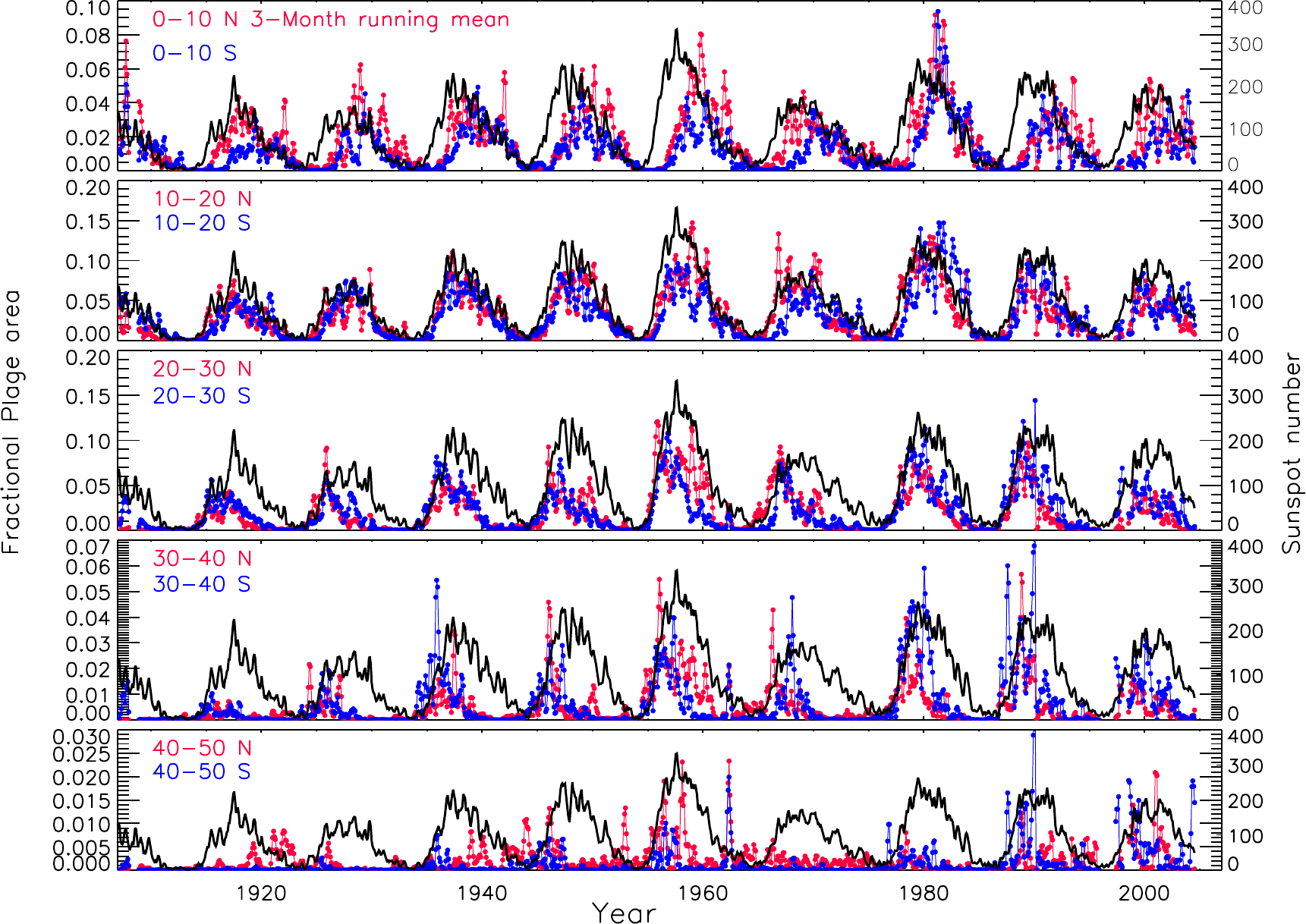}
\caption{Five panels of the figure show the variation of fractional plage area on monthly with 3-months running average basis for the period 1907–2007 for the equatorial belts (5$^{\circ}$ – 45$^{\circ}$) at an interval of 10$^{\circ}$ for the northern (red) and southern (blue) hemispheres. Sunspot data with the same averages but for the whole Sun is overplotted for comparison. The latitude belt has been indicated in each panel. }
\label{fig:4}
\end{figure}

To study the variation of plage area with time, we have computed the fractional area in each latitude belt, considering the area of that latitude belt. Monthly averages are computed using the daily data. Then these values were smoothed by taking the running average over three months to study the long-term variations. In 5 panels of Figure~\ref{fig:4}, we show fractional plage area on a monthly basis for 0$^{\circ}$ -- 10$^{\circ}$, 10$^{\circ}$ -- 20$^{\circ}$, 20$^{\circ}$ -- 30$^{\circ}$, 30$^{\circ}$ -- 40$^{\circ}$, and 40$^{\circ}$ -- 50$^{\circ}$ northern latitude belts in red and in blue for southern belts for the period of 1907 -- 2007. The latitude belt is indicated in each panel. Hereafter, we refer to these by the mean latitude as 5$^{\circ}$, 15$^{\circ}$, 25$^{\circ}$, 35$^{\circ}$, and 45$^{\circ}$ latitude belts. The sunspot number representing the whole solar disk with the same averaging procedure is shown in black color. The comparison of the variations in Ca-K plage area at 25$^{\circ}$ and 35$^{\circ}$ latitude belts with sunspot number on the disc indicates that maximum Ca-K plage area occurs at these belts before the sunspot maximum whereas at 5$^{\circ}$ belt, it happens after the sunspot maximum. It appears that the phase difference between sunspot numbers and fractional plage area at the 15$^{\circ}$ belts is minimum. We have plotted the fractional plage area up to 50$^{\circ}$ latitude belts, as these are rarely seen at higher latitudes. The variations in the small-scale Ca-K features such as EN, AN, and QN are similar to those of plages. The details of variations in small-scale features are shown in Appendix A.

\subsection{Phase difference between activities at different latitude belts}

\begin{figure}[!ht]
\centering
\includegraphics[width=0.75\textwidth]{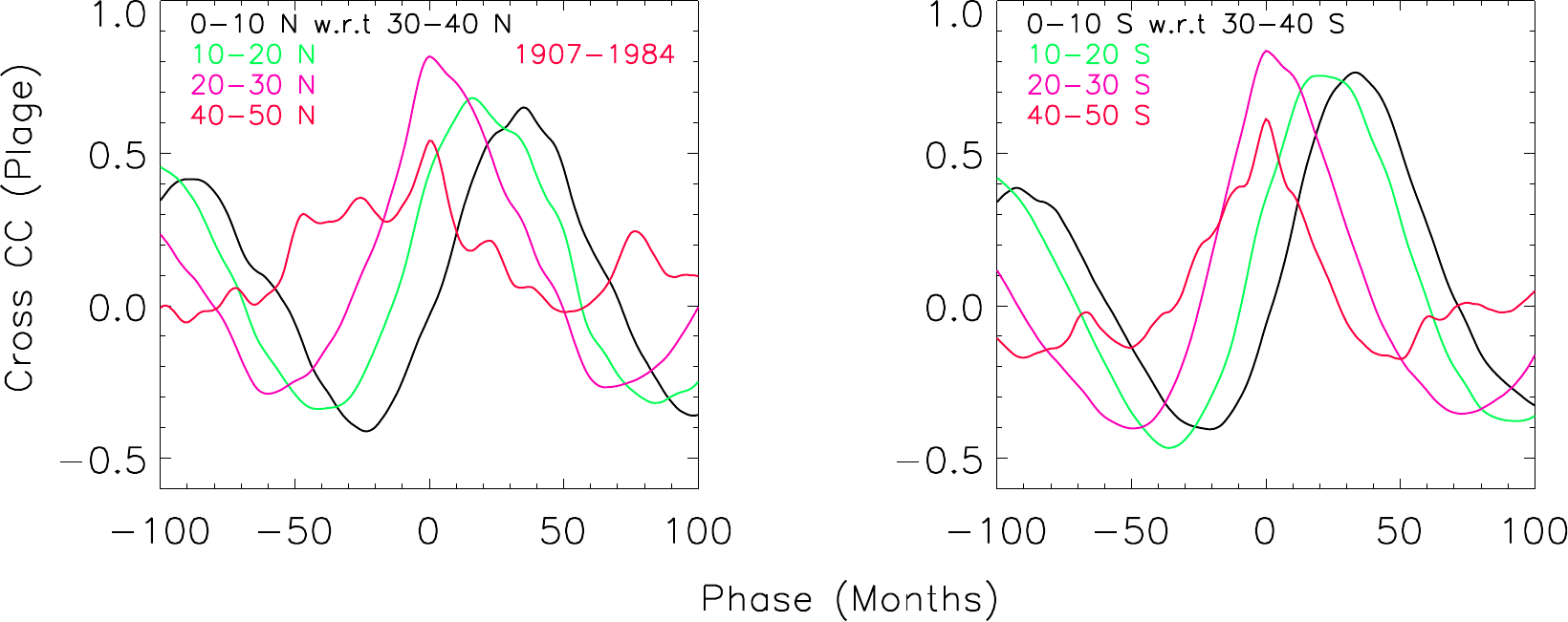}
\caption{Cross correlation curves for 5$^{\circ}$(black), 15$^{\circ}$(green), 25$^{\circ}$(pink) and 45$^{\circ}$(red) with respect to 35$^{\circ}$ latitude for Ca-K plage area considering the period of 1907 – 1984. }
\label{fig:5}
\end{figure}

To determine the speed of activity shift in the Sun as a function of latitude, we computed phase difference using monthly data. We have used a cross-correlation analysis to determine the phase shift in the activity between the latitudes in steps of one month from ${\pm}$1 to 100 months. Generally, the active regions first appear around 35$^{\circ}$ latitudes at the beginning of a solar cycle. As the cycle progresses, the plages (representing the toroidal magnetic field) gradually appear at lower latitudes. Therefore, to estimate the speed of the shift, we have computed the phase difference for various latitude belts with the 35$^{\circ}$ belt. We have used the monthly fractional area with a running average over 13 months to determine the phase difference. Two panels of Figure \ref{fig:5} show the cross-correlation values of 5$^{\circ}$, 15$^{\circ}$, 25$^{\circ}$, and 45$^{\circ}$ latitude belts with respect to 35$^{\circ}$ belt for the fractional plage area for southern and northern hemispheres considering the data for the period 1907 – 1984 together. The data for the period 1985 -- 2007 has not been considered because of significant gaps in the observations. 

\begin{figure}[!ht]
\centering
\includegraphics[width=0.75\textwidth]{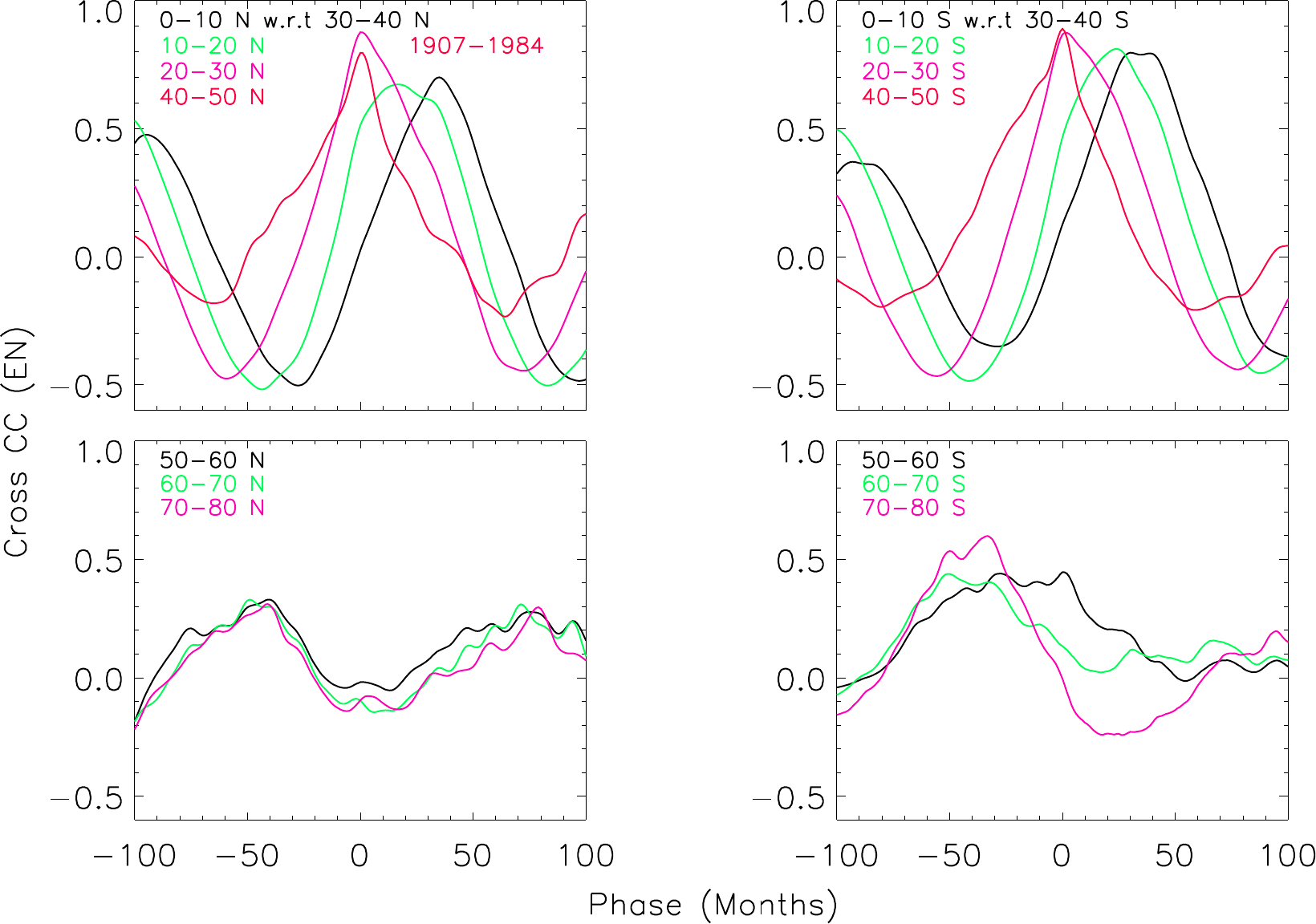}
\caption{Cross correlation curves for various latitude belts with respect to 35$^{\circ}$ latitude belt for EN area considering the data for the period of 1907 – 84.}
\label{fig:6}
\end{figure}

 Two panels in the upper row of Figure \ref{fig:6} show the cross-correlation function for 5$^{\circ}$, 15$^{\circ}$, 25$^{\circ}$, and 45$^{\circ}$ belts with respect to 35$^{\circ}$ belt for the fractional EN area for both the hemispheres. The two panels in the bottom row indicate the cross-correlation function for 55$^{\circ}$, 65$^{\circ}$, and 75$^{\circ}$ belts with 35$^{\circ}$ belt. We have also computed the cross-correlation values for fractional AN and QN for all the belts. The results for AN and QN are similar to those for EN and plage areas.  We have also computed the phase difference between the activity at different latitudes for individual solar cycles, the results are discussed in Appendix B.
 
 \par The occurrences of peak and valley of small scale structures (networks) in the time series follow the variation in large scale features such as sunspots and plages. The plage areas represent the large scale toroidal field and networks refer to poloidal fields. Hereafter, we refer this temporal variation in plage and network areas, as ``activity''. It may be noted that plages occur up to ${\sim}$ 50$^{\circ}$ only whereas networks are visible on the whole solar surface. While discussing the polar region, we imply small scale features (networks) only. The phase difference between the two latitude belts gives the time difference between the occurrences of maximum area (peak) of activity at the two latitude belts. The values of the amplitude of cross correlation coefficient and phase difference in months for various latitude belts with respect to 35$^{\circ}$ latitude are listed in Table \ref{tab:1}. The negative phase (Table \ref{tab:1}) difference between 45$^{\circ}$ and 35$^{\circ}$ latitude belts implies that the maximum area of networks occurred at 45$^{\circ}$ belt occurred earlier than that at 35$^{\circ}$ belt. The values of the phase difference for different latitude belts imply that the shift-speed of the activity at the beginning of the solar cycle is more than near the end of the cycle. The difference in time for the occurrence of maximum activity at 35$^{\circ}$ and 5$^{\circ}$ is about 36 months in both hemispheres.

\begin{table}[!ht]
\centering
\caption{The values of CC and phase difference for plage, EN, AN, and QN for various latitude belts with respect to 35$^{\circ}$ belt considering the period of 1907--84. ``NSP'' stands for when no significant peak is found in the cross-correlation curve.}
\label{tab:1}
\vspace*{5mm}
\scalebox{0.7}{%
\begin{tabular}{ |c|c|c|c|c|c|c|c|c| }
 \hline
\textbf{Cross Correlation}  & \multicolumn{4}{c|}{\textbf{Maximum value - Cross correlation}}  & \multicolumn{4}{c|}{\textbf{Phase difference in months}} \\
\textbf{ between latitudes}  & \multicolumn{4}{c|}{\textbf{coefficients}} & \multicolumn{4}{c|}{\textbf{}} \\
 \hline
   & \textbf{Plage}  &	\textbf{EN} &	\textbf{AN} &	\textbf{QN} & \textbf{Plage}  & \textbf{EN} & \textbf{AN} & \textbf{QN} \\ 
\hline

05$^{\circ}$ -- 35$^{\circ}$ N&0.65&0.7&0.70&0.46&37&36&35&35\\
\hline
15$^{\circ}$ -- 35$^{\circ}$ N&0.86&0.67&0.74&0.67&18&22&21&19\\
\hline
25$^{\circ}$ -- 35$^{\circ}$ N&0.82&0.88&0.90&0.90&7&10&9&5\\
\hline
35$^{\circ}$ -- 35$^{\circ}$ N&1&1&1&1&0&0&0&0\\
\hline
45$^{\circ}$ -- 35$^{\circ}$ N&0.54&0.80&0.70&0.76 &--4 &- 6 & --5 &--4\\
\hline
55$^{\circ}$ -- 35$^{\circ}$ N&NSP&0.33,0.28&0.33,0.35&NSP&NSP&- 51, 72 &--54, 70 & NSP\\
\hline
65$^{\circ}$ -- 35$^{\circ}$ N&NSP&0.32,0.33&0.29,31&NSP&NSP& --53, 63 & --56, 67 & NSP\\
\hline
75$^{\circ}$ -- 35$^{\circ}$ N&NSP&0.30,0.32&0.33,31&NSP&NSP& --50, 68 & --50, 72 & NSP\\
\hline
05$^{\circ}$ -- 35$^{\circ}$ S&0.76&0.80&0.81&0.64&34&35&37&41\\
\hline
15$^{\circ}$ -- 35$^{\circ}$ S&0.75&0.81&0.79&0.68&19&23&24&21\\
\hline
25$^{\circ}$ -- 35$^{\circ}$ S&0.85&0.87&0.88&0.84&6&9&8&5\\
\hline
35$^{\circ}$ -- 35$^{\circ}$ S&1&1&1&1&0&0&0&0\\
\hline
45$^{\circ}$ -- 35$^{\circ}$ S&0.61&0.89&0.82&0.81& --5&  --6 &  --3 & --2\\
\hline
55$^{\circ}$ -- 35$^{\circ}$ S&NSP&0.45&0.32,0.28&0.54,0.22&NSP&  --34& --42, 7& --33, 73\\
\hline
65$^{\circ}$ -- 35$^{\circ}$ S&NSP&0.44&0.47,0.34&0.45,0.24&NSP& --46 & --38, 75& --35, 74\\
\hline
75$^{\circ}$ -- 35$^{\circ}$ S&NSP&0.60&0.45,0.26&0.44, NSP&NSP& --42 & --45, 71 & --34, NSP\\
\hline
\end{tabular}}
\end{table}

\subsection{Phase difference between activities at various latitude belts and Sunspot number}

\begin{figure}[!ht]
\centering
\includegraphics[width=0.75\textwidth]{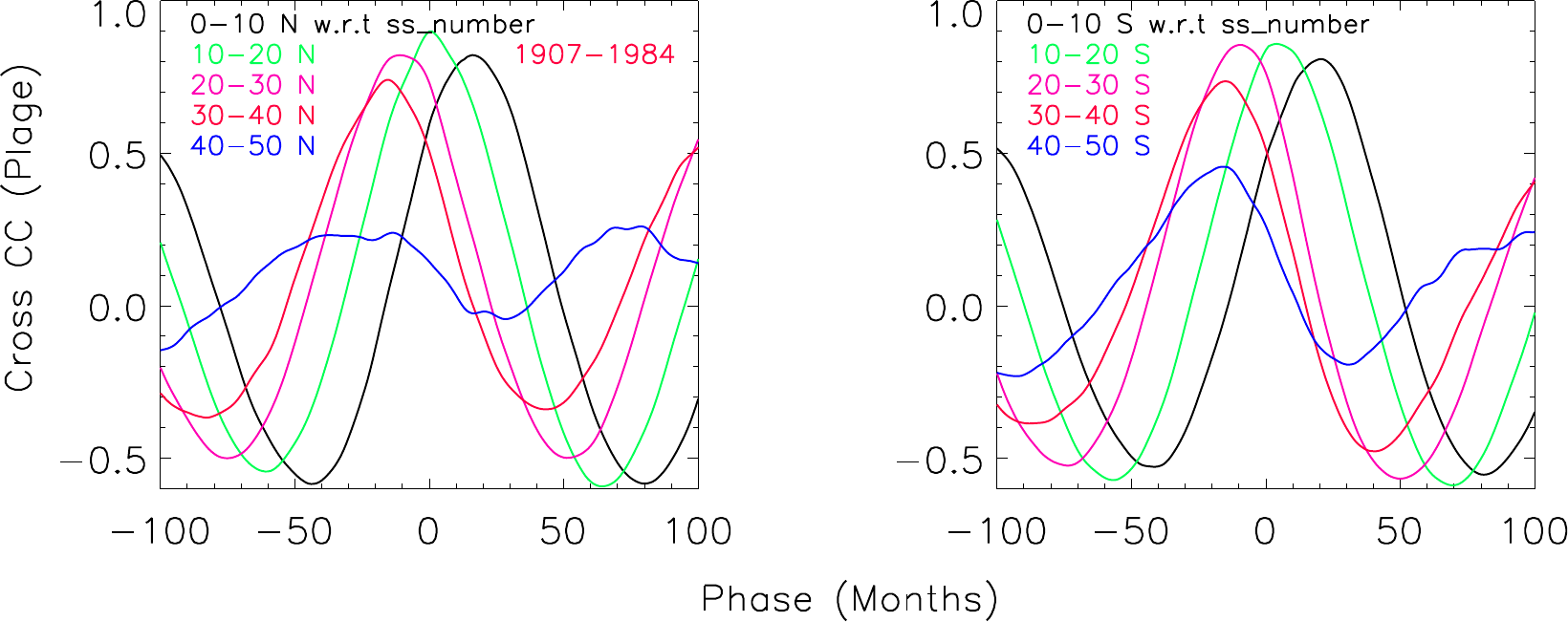}
\caption{Cross correlation curves of Ca-K plage area for 5$^{\circ}$, 15$^{\circ}$, 25$^{\circ}$, 35$^{\circ}$, and 45$^{\circ}$ latitude belts with respect to sunspot number (SS) over the visible disk considering the period of 1907 – 1984.}
\label{fig:7}
\end{figure}

\begin{figure}[!ht]
\centering
\includegraphics[width=0.75\textwidth]{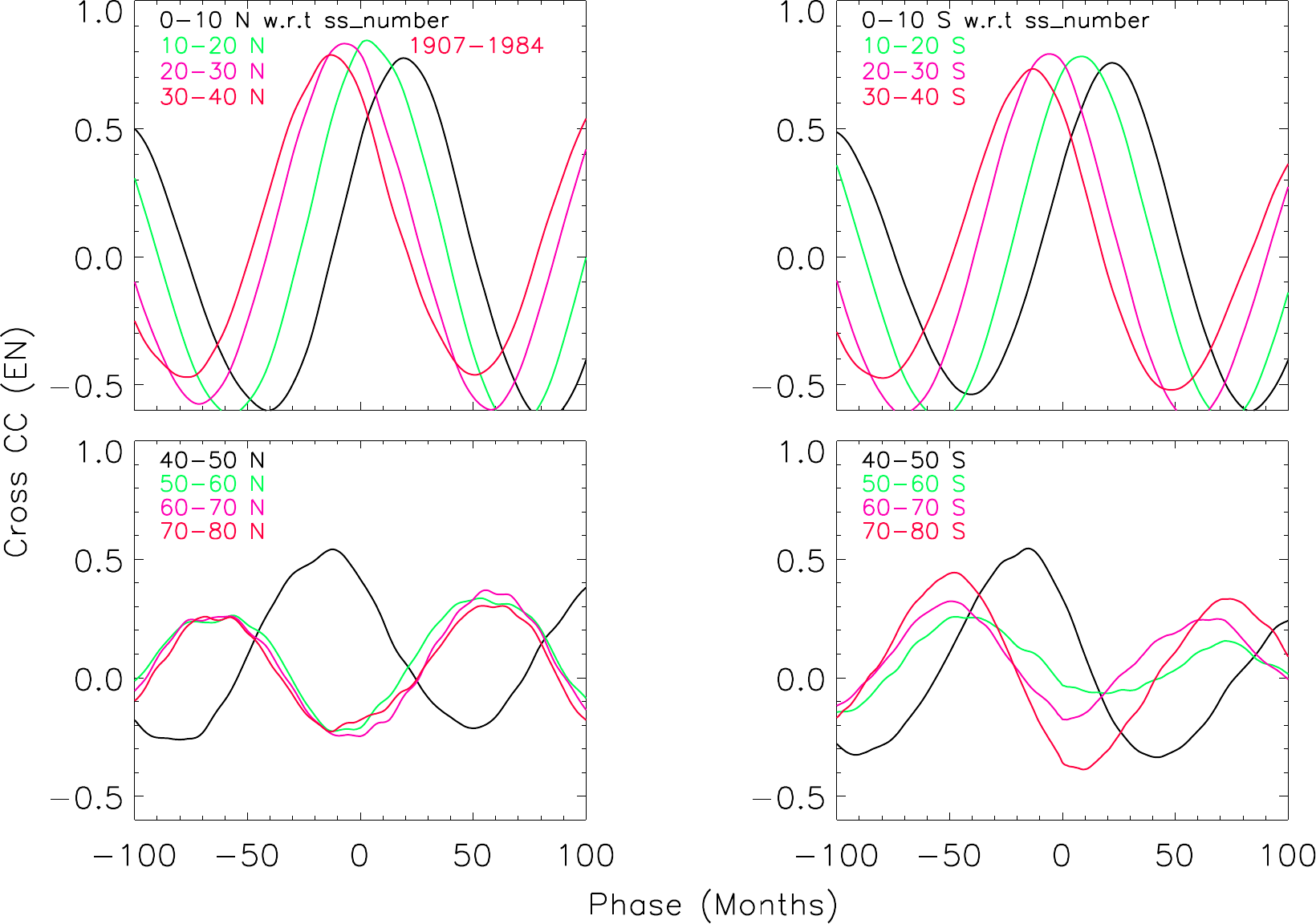}
\caption{Cross correlation curves of Ca-K EN area for 5$^{\circ}$, 15$^{\circ}$, 25$^{\circ}$, 35$^{\circ}$, and 45$^{\circ}$ latitude belts with respect to sunspot number (SS) over the visible disk considering the period of 1907 – 1984.}
\label{fig:8}
\end{figure}

The sunspot number is a good indicator of solar activity. For each cycle, the sunspot number attains a maximum value at a certain epoch that can be used as a reference to compute the phase difference between sunspot and Ca-K activities at different latitudes. We have computed the cross-correlation functions for the Ca-K features such as plage, EN, AN, and QN area with respect to sunspot numbers (SS) over the visible disk. The cross-correlation curves for the plage area are shown in Figure~\ref{fig:7} and those for the EN area in Figure~\ref{fig:8}.  The values of phase differences between the occurrences of the area of Ca-K features at various latitudes and SS considering the data for the period 1907--84 are listed in Table~\ref{tab:2}. The correlation functions appear better and more consistent with the SS data compared to the Ca-K features at 35$^{\circ}$ latitude belt. The correlation functions for 55$^{\circ}$ -- 75$^{\circ}$ latitude belts are similar, showing almost the same phase difference with respect to SS for the high latitude belts. The phase differences for individual cycles for EN and AN areas are given in Tables~\ref{tab:C1} and \ref{tab:C2} (Appendix C) for the northern and southern hemispheres.

\begin{table}[!ht]
\centering
\caption{The values of phase differences for plage, EN, AN, and QN for various latitude belts with respect to sunspot number (SS) over the visible solar disk considering the period of 1907–84 are listed. ``NSP'' stands for when no significant peak is found in the cross-correlation curve.}
\label{tab:2}
\vspace*{5mm}
\scalebox{0.7}{%
\begin{tabular}{ |c|c|c|c|c|c|c|c|c| }
 \hline
\textbf{Cross Correlation between }  & \multicolumn{4}{c|}{\textbf{Phase difference in months}}  & \multicolumn{4}{c|}{\textbf{Phase difference in months}} \\
\textbf{latitudes with SS}  & \multicolumn{4}{c|}{\textbf{for northern hemisphere}}  & \multicolumn{4}{c|}{\textbf{for southern hemisphere}} \\
 \hline
   & \textbf{Plage}  &	\textbf{EN} &	\textbf{AN} &	\textbf{QN} & \textbf{Plage}  & \textbf{EN} & \textbf{AN} & \textbf{QN} \\ 
\hline
05$^{\circ}$  – SS& 18& 20& 18& 16    & 20& 21& 22& 21   \\
\hline

15$^{\circ}$ – SS& 1& 3 & 3&  1     & 3& 7& 5  & 5     \\
\hline
25$^{\circ}$ – SS& - 9& - 7&- 6&- 8    &- 9&- 7&- 8&- 7   \\
\hline

35$^{\circ}$ – SS&- 14&- 12&- 13&- 13   &- 15&- 13&- 14&- 14  \\
\hline
45$^{\circ}$ – SS&- 20&- 15&- 20&- 18   &- 18&- 16&- 19&- 17 \\
\hline

55$^{\circ}$ – SS&NSP&- 62 , 54 &- 65 , 56&- 65, 56&NSP&- 45 , 73&- 51 , 68&- 59 , 70\\
\hline

65$^{\circ}$  – SS&NSP&- 64 , 56&- 67, 60&- 63, 59&NSP&-49, 70&- 59 , 63&- 51 , 65\\
\hline

75$^{\circ}$  – SS&NSP&- 65 , 52&- 64 , 62&- 60, 64&NSP&- 48 , 75&- 54 , 70&- 53 , 71\\

\hline
\end{tabular}}

\end{table}

\section{Discussions}

We have computed the cross-correlation values between the activity at different latitude belts and 35$^{\circ}$ belt up to 100 months on both sides of zero phase difference. The phase differences for the Ca-K line features such as plage, EN, AN, and QN with respect to corresponding activity at the 35$^{\circ}$ latitude belt show a similar trend for all the features and for all the solar cycles. The values of phase differences vary for each cycle, but it is difficult to say if these are of solar origin or within statistical limits. The Table \ref{tab:1} shows positive phase difference for the 25$^{\circ}$, 15$^{\circ}$, and 5$^{\circ}$ latitude belts with respect to 35$^{\circ}$ belts whereas negative phase difference between 35$^{\circ}$ and 45$^{\circ}$ latitude belt. This implies that activity begins around the 45$^{\circ}$ latitude belt and then shifts to lower latitude belts. The average values of phase differences indicate that it takes about 5, 8, 12, and 16 months for the activity to travel from 45$^{\circ}$ to 35$^{\circ}$, 35$^{\circ}$ to 25$^{\circ}$, 25$^{\circ}$ to 15$^{\circ}$, and 15$^{\circ}$ to 5$^{\circ}$, respectively. This implies that activity travel-speed is ${\sim}$ 9.6 m/sec at the beginning of the solar cycle at ${\sim}$ 45$^{\circ}$ latitude and decreases to ${\sim}$ 3 m/sec near the end of the cycle at ${\sim}$ 5$^{\circ}$ latitude. The phase difference for 55$^{\circ}$, 65$^{\circ}$, and 75$^{\circ}$ belts with respect to 35$^{\circ}$ latitude belts remains the same.

\par We have also computed the cross-correlation of Ca-K features at different latitudes with respect to sunspot numbers (SS) on the visible disk. The average phase differences for all the features considering the data of 1907 – 84 indicate that it takes about 5, 6, 9, and 16 months for the activity to travel from 45$^{\circ}$ to 35$^{\circ}$, 35$^{\circ}$ to 25$^{\circ}$, 25$^{\circ}$ to 15$^{\circ}$, and 15$^{\circ}$ to 5$^{\circ}$, respectively, in the northern hemisphere whereas it takes 4, 6, 13, and 16 months in the southern hemisphere. Comparatively, the activity took three more months for the southern as compared to the northern hemisphere to travel from 45$^{\circ}$ to 5$^{\circ}$ latitude. The cross-correlation functions (Table \ref{tab:2}) for the Ca-K features at the 65$^{\circ}$ and 75$^{\circ}$ belts with the sunspot numbers show two peaks, one with a phase difference of ${\sim}$ -5 and the other at +5 years. Both these peaks are separated by 10 -- 11 years. Table \ref{tab:2} also shows no systematic phase difference between the activity at 55$^{\circ}$, 65$^{\circ}$, and 75$^{\circ}$ latitudes. 

\par  The toroidal fields in the form of sunspots and plages appear at the solar surface around mid-latitudes at the beginning of the solar cycle. They then occur progressively at lower and lower latitudes as the cycle progresses \citep{babcock1961}. The poloidal component of the magnetic field moves towards the polar region with time. It may create variation in small-scale activity at the polar latitude belt with time caused by the dynamo process operating at the base of the convection zone of the Sun and meridional flows. There is a phase difference between the mid and lower latitudes and also between the mid and polar latitudes. Also, there are minimum variations in the network areas at ${\sim}$ 60$^{\circ}$ latitude with the solar cycle as compared to other latitudes, including polar region. These findings, suggest that there could be two meridional circulation cells in the latitudinal direction as suggested by \citet{dikpati2010, dikpati2012} and \citet{bernadett2015}. The one cell of primary meridional flow is circulating from the equator up to 60$^{\circ}$ latitude. The second cell has counter circulation from poles towards 60$^{\circ}$ latitude. This makes the path shorter for magnetic flux transport through meridional circulation. The shorter path produces the smaller cycle with less than 11-year period. On the other hand, the \citet{babcock1961} and \citet{leighton1969} dynamo models suggest one meridional cell which runs from equator to poles and return flow in the deep convection zone that can produce such effect. It may be noted that the occurrence of maximum area of networks in the polar region is anti-correlated with that of sunspot maxima (toroidal field) during a solar cycle as reported by \citet{makarov2004}. Also, there is no significant phase difference between the large-scale and small-scale activity at the middle and lower latitude belts. 

\section{Summary}

Earlier \citet{singh2021, singh2022} showed that the application of ECT to the Ca-K images makes the data uniform and permits to study of long-term variations in small-scale features reliably. Here, we demonstrated that plage, EN, AN, and QN area could be used to study the variation of these features as a function of latitude and time that has a direct bearing on the meridional flows generated by the dynamo at the base of the convection zone and just below the solar surface. We find that large-scale active regions appear at Sun's surface at ${\sim}$ 45$^{\circ}$ latitude at the beginning of the solar cycle. The small-scale active areas also increase as the Ca-K plage areas increase at respective latitude belts. The average value of phase difference of 5 months between 45$^{\circ}$ and 35$^{\circ}$ (Table \ref{tab:1}) indicate that meridional flow has a velocity of about 9.4 m/sec at the beginning of the solar cycle at ${\sim}$40$^{\circ}$ latitude belt. Considering the average phase difference (all networks) for the other latitude belts we find that speed decreases to ${\sim}$ 6, ${\sim}$ 4.5 and ${\sim}$ 3~m/sec at 25, 15, and 5$^{\circ}$ latitude belts. These are average values inferred from phase differences in the occurrence of activity at these latitude belts. The cross-correlation function and phase differences indicate that the activity in the polar region is anti-correlated with the sunspot number. There is no phase difference between the activity at 55$^{\circ}$, 65$^{\circ}$, and 75$^{\circ}$ latitude belts raising the doubt that the poloidal field moves from middle latitude belts to the polar region. These findings point towards the existence of multi-cells in the convection zone.

\section{Acknowledgements}

We thank the reviewer for the valuable comments. We thank the numerous observers who kept the data in good shape and the digitization team for doing the laborious work. The digitization of the Kodaikanal solar data was planned by Jagdev Singh and fabrication of the digitizers was supported by F. Gabriel and P U. Kamath.

\appendix

\section{Variations of small scale Ca-K features with time}

 The variation of plage area has been studied by many over long periods, but variation in small-scale Ca-K features (EN, AN, and QN) is investigated significantly less due to the unavailability of properly calibrated images and due to the change in contrast of the images on a short and long term basis as mentioned in the introduction. To study the long term variation in Ca-K features, we do a monthly average of the data to reduce the effect of rotational modulation of the sun. We have again done a 3-month running average to suppress the remnant effects of rotation modulation.
 
 Figure~\ref{fig:A1} shows the variation of fractional EN area as a function of time on a monthly basis with a running average over three months for 45$^{\circ}$ south to 45$^{\circ}$ north latitudes at an interval of 10$^{\circ}$ for the period 1907 -- 2007. A large amplitude variations in the fractional EN area occur around 15$^{\circ}$ latitude belts with the phase of the solar cycle as compared to other latitude belts. Also, the variations at 15$^{\circ}$ latitude belts appear to be in phase with the sunspot cycle. The behavior of EN is similar to that of plage area. Three panels of Figure \ref{fig:A2} show the fractional EN area for 55$^{\circ}$ -- 75$^{\circ}$ latitude belts for both the hemispheres in red (north) and blue (south) at an interval of 10$^{\circ}$ as a function time on a monthly basis with a running average of 13 months. Figures \ref{fig:A1} and \ref{fig:A2} indicate that fractional EN area at all latitude belts shows an 11-year periodicity. 

Figures \ref{fig:A3} and \ref{fig:A4} show the fractional AN area variation at various latitude belts as a function of time with the same parameters as those of the EN area. And Figures \ref{fig:A5} and \ref{fig:A6} indicate the fractional QN variations on a monthly basis with the 13-month running average. All figures \ref{fig:A1} to \ref{fig:A6} show similar variations in EN, AN, and QN with varying amplitude as a function of feature and latitude. The small-scale features show 11-year solar cycle variations up to 75$^{\circ}$ latitude belts. Figures \ref{fig:A2} for EN, \ref{fig:A4} for AN and \ref{fig:A6} for QN area indicate that occurrence of small scale features in polar region (latitude $>$ 65 $^{\circ}$) is anti-correlated with the sunspot numbers.

\setcounter{figure}{0}
\renewcommand{\thefigure}{A{1}}

\begin{figure}[!ht]
\centering
\includegraphics[width=0.8\textwidth]{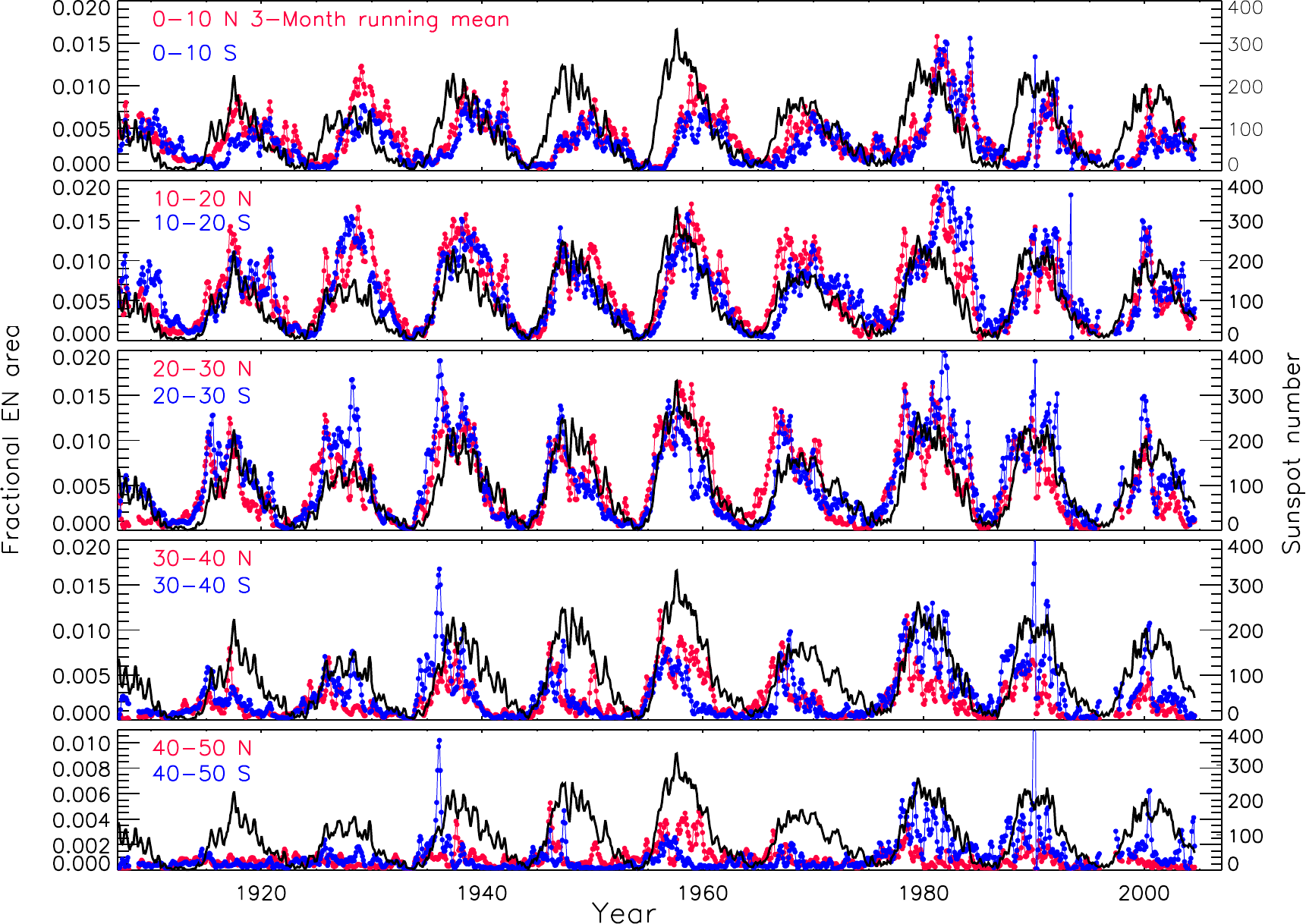}
\caption{Five panels of the figure show the variation of fractional plage area on monthly with 3-months running average basis for the period 1907 -- 2007 for the equatorial belts (5$^{\circ}$ -- 45$^{\circ}$) at an interval of 10$^{\circ}$ for the northern (red) and southern (blue) hemispheres. Sunspot data with the same averages but for the whole Sun is over plotted for comparison. The latitude belt has been indicated in each panel.}
\label{fig:A1}
\end{figure}

\renewcommand{\thefigure}{A{2}}
\begin{figure}[!ht]
\centering
\includegraphics[width=0.8\textwidth]{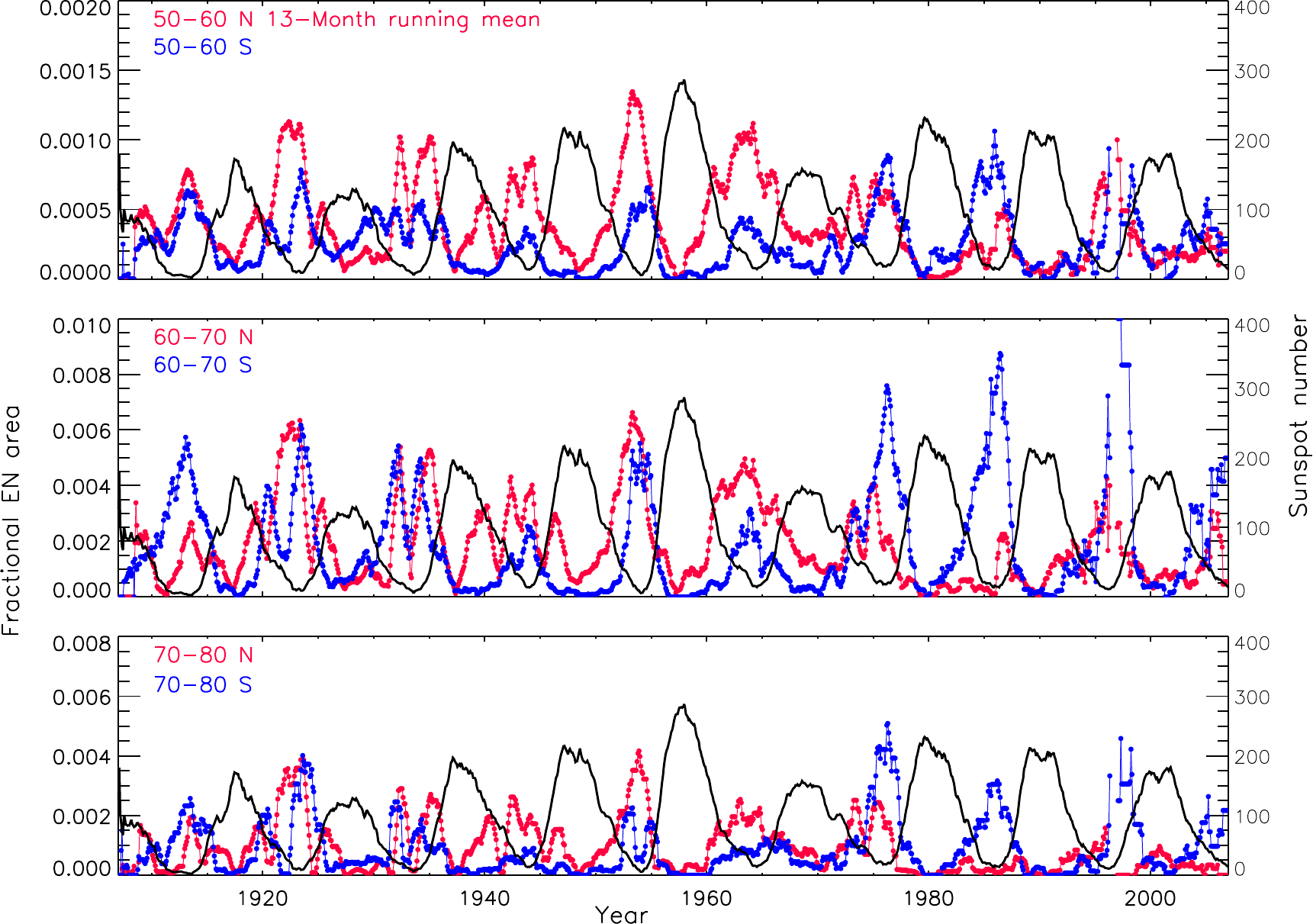}
\caption{Three panels of the figure show the variation of fractional EN area with 13-months running average for the period 1907 -- 2007 for the polar latitude belts (55$^{\circ}$ -- 75$^{\circ}$) at an interval of 10$^{\circ}$ for the northern (red) and southern (blue) hemispheres. Sunspot data with the same averages but for the whole Sun is overplotted for comparison. The latitude belt has been indicated in each panel. }
\label{fig:A2}
\end{figure}

\renewcommand{\thefigure}{A{3}}
\begin{figure}[!ht]
\centering
\includegraphics[width=0.8\textwidth]{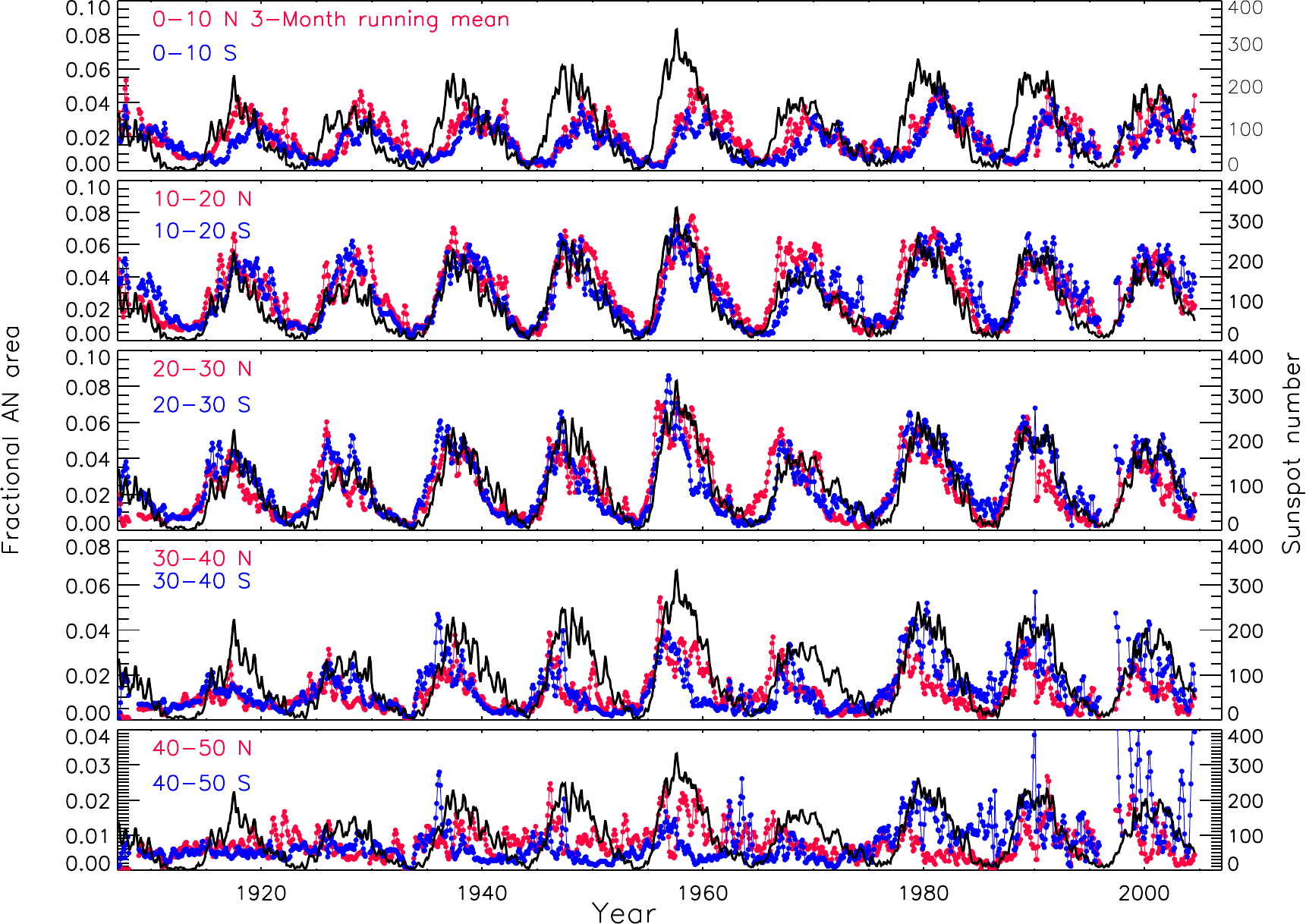}
\caption{Same as that for Figure \ref{fig:A1} but for fractional AN area.}
\label{fig:A3}
\end{figure}

\renewcommand{\thefigure}{A{4}}
\begin{figure}[!ht]
\centering
\includegraphics[width=0.8\textwidth]{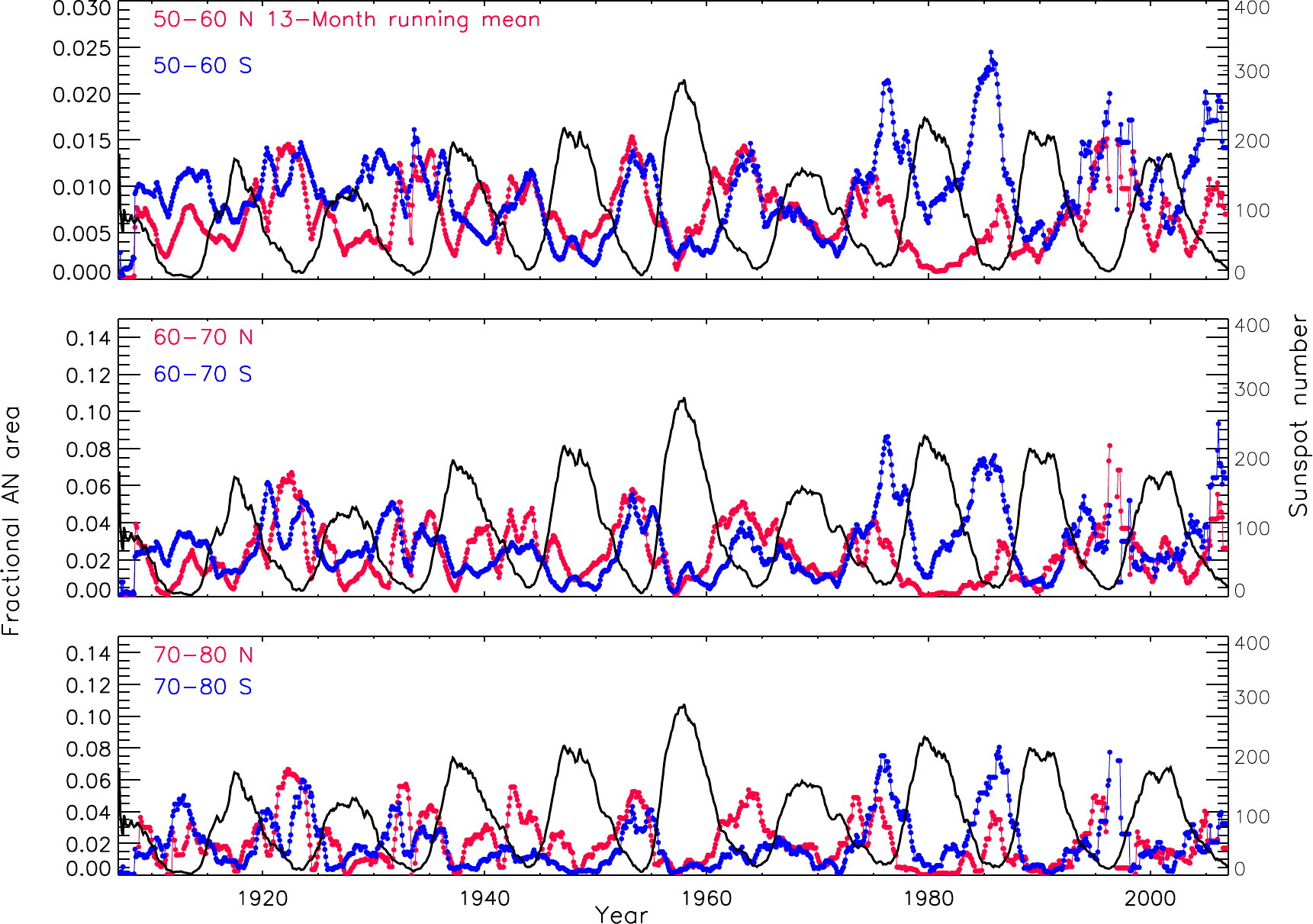}
\caption{Same as that for Figure \ref{fig:A2} but for fractional AN area. }
\label{fig:A4}
\end{figure}

\renewcommand{\thefigure}{A{5}}
\begin{figure}[!ht]
\centering
\includegraphics[width=0.8\textwidth]{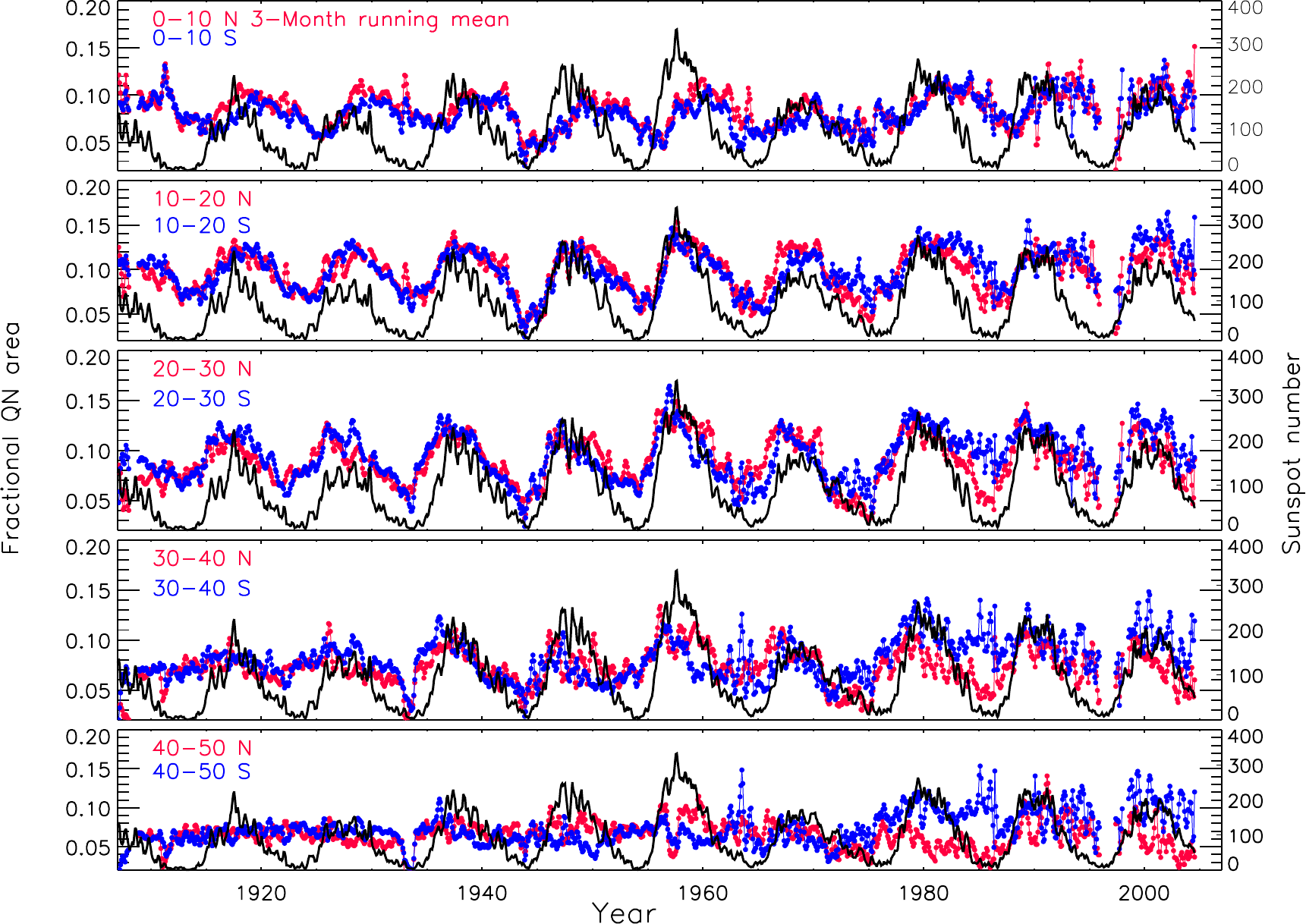}
\caption{Same as that for Figure \ref{fig:A1} but for fractional QN area. }
\label{fig:A5}
\end{figure}

\renewcommand{\thefigure}{A{6}}
\begin{figure}[!ht]
\centering
\includegraphics[width=0.8\textwidth]{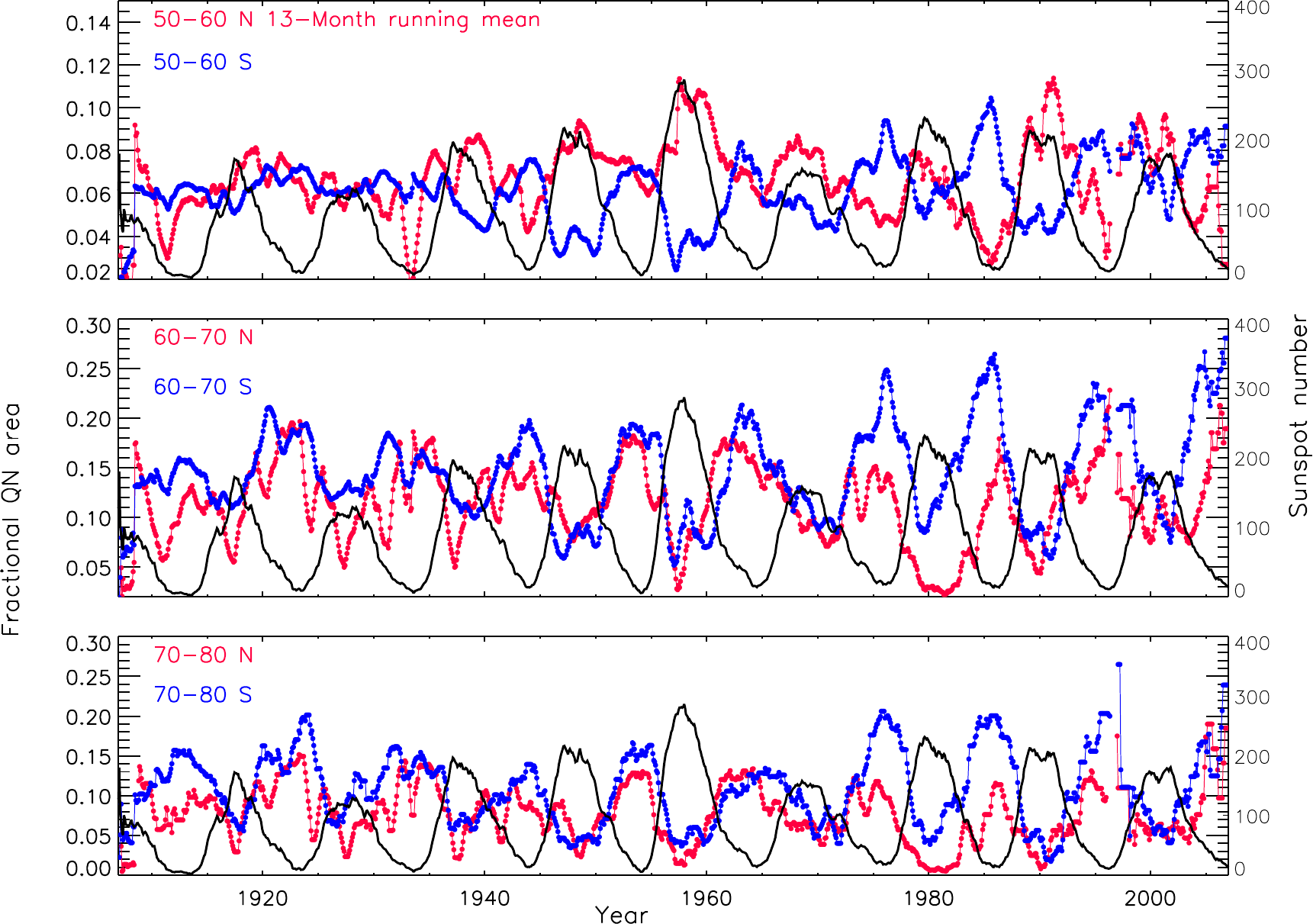}
\caption{Same as that for Figure \ref{fig:A2} but for fractional QN area. }
\label{fig:A6}
\end{figure}

\section{Phase difference between activity at different latitudes for individual solar cycle}

We have also computed the phase difference in the activity between latitude belts for each solar cycle to determine its relationship with the strength of the solar cycle. To determine the speed of shift in activity for the individual cycle, we have selected the data of 16 years centered around the maximum phase of each cycle as listed in Table~\ref{tab:B1}. The cross-correlation values for each data set representing cycle numbers 15 to 21 were computed. Figure~\ref{fig:B1} shows the cross-correlation curves between different latitude belts (5$^{\circ}$ to 75$^{\circ}$) and 35$^{\circ}$ belt for the EN area of solar cycle number 19 for both the hemispheres. We list the phase difference values for the EN and AN for various latitudes in the northern hemisphere in Table~\ref{tab:B1} and the southern hemisphere in Table \ref{tab:B2} for individual cycles. The `NSP' in the Tables indicates no significant peak in the respective cross-correlation curve. The phase differences for plage area are computed only up to 45$^{\circ}$  and for other features up to 75$^{\circ}$. The values indicate that activity shifts from the middle latitude belt to lower latitude belts at a faster rate at the beginning of the cycle compared to near the end of the solar cycle for all the analyzed data.

\setcounter{figure}{0}
\renewcommand{\thefigure}{B{1}}

\begin{figure}[!ht]
\centering
\includegraphics[width=0.75\textwidth]{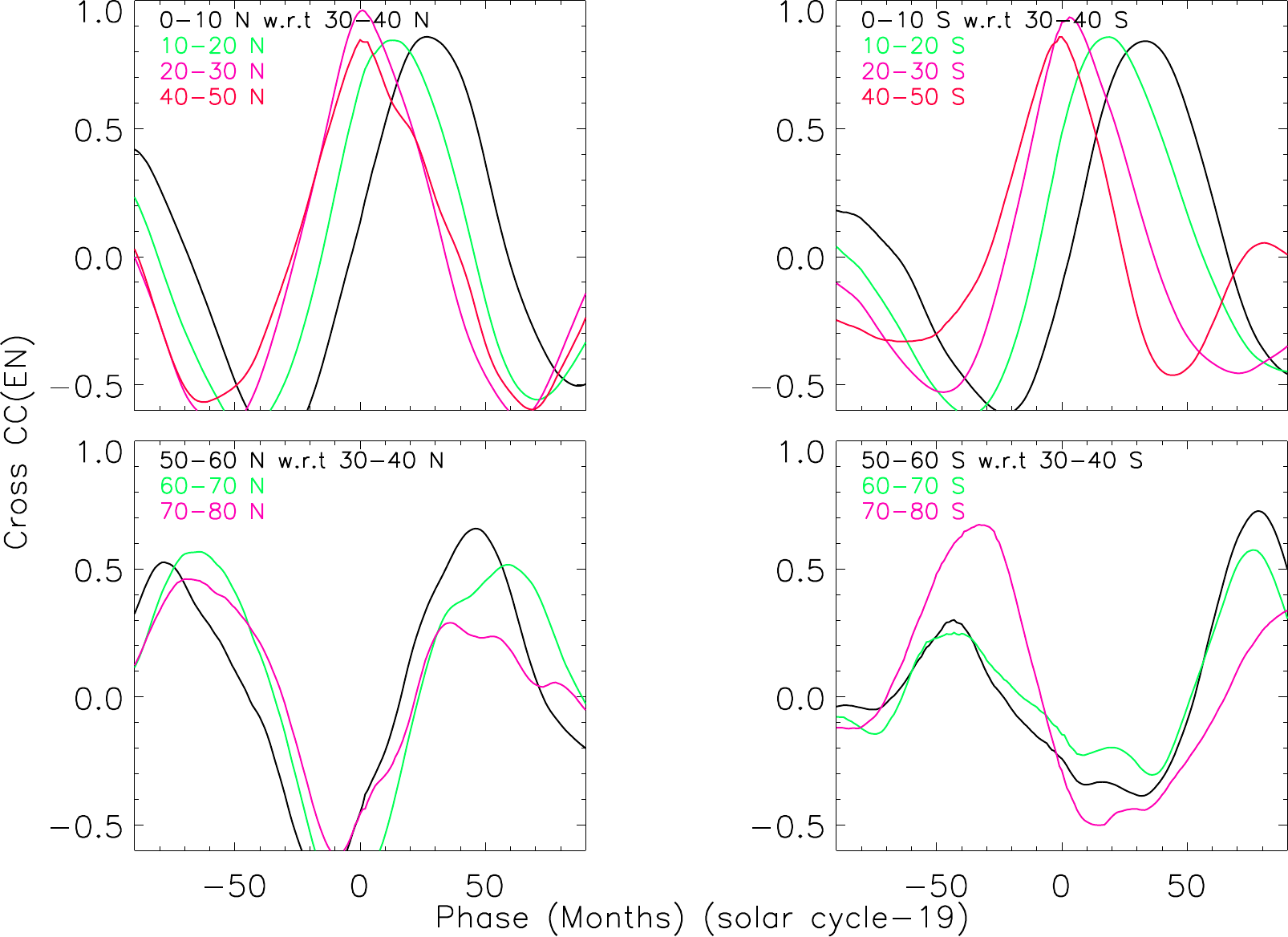}
\caption{Cross correlation curves for various latitude belts with respect to 35$^{\circ}$ latitude belt for EN area for the solar cycle 19 considering the data for the period of 1952 -- 67.}
\label{fig:B1}
\end{figure}

\setcounter{table}{0}
\renewcommand{\thetable}{B{1}}

\begin{table}[!ht]
\centering
\caption{Lists the phase difference in months between activity at various latitude belts with respect to 35$^{\circ}$ belt for the northern hemisphere. ``NSP'' stands for when no significant peak is found in the cross-correlation curve.}
\label{tab:B1}
\vspace*{5mm}
\scalebox{0.7}{%
\begin{tabular}{ |c|c|c|c|c|c|c|c|c| }
 \hline
   \textbf{Cycle number} & \textbf{Period} &  \textbf{5$^{\circ}$} & \textbf{15$^{\circ}$} & \textbf{25$^{\circ}$} & \textbf{45$^{\circ}$} & \textbf{55$^{\circ}$} & \textbf{65$^{\circ}$} & \textbf{75$^{\circ}$}\\
\hline
                     \multicolumn{9}{|c|}{\textbf{EN area}} \\
\hline
15&1910 - 25&36&17&4&- 4&NSP, 56&NSP,58&NSP , 58 \\
\hline
16&1921 - 36&42&22&2&- 6&- 32 , NSP&- 34 , NSP&- 42 , NSP \\
\hline
17&1931 - 46&29&11&3&- 2&NSP&NSP&NSP \\
\hline
18&1941 - 56&36&20&2&- 2&- 34 , 82&- 38 , 80&- 40 , 81 \\
\hline
19&1952 - 67&28&14&2&0&- 78 , 46&- 65 , 50&- 63 , 55 \\
\hline
20&1962 - 77&31&16&4&- 3&- 47 , 72&- 53 , 69&- 44 , 75 \\
\hline
21&1973 - 84&37&18&4&- 5&- 55 , 75&- 56 , 77&- 58 , 70 \\
\hline
                   \multicolumn{9}{|c|}{\textbf{AN area}}\\
\hline
15&1910 - 25&31&17&4&NSP&- 38 , 57&- 32 , 60&- 24 , 65 \\
\hline
16&1921 - 36&41&23&2&- 5&- 47 , NSP&- 47 , NSP&- 43 , NSP \\
\hline
17&1931 - 46&31 &11&2&0&NSP&NSP&NSP \\
\hline
18&1941 - 56&32&14&2&0&- 38 , 82&- 40 , 81&- 36 , 83 \\
\hline
19&1952 - 67&29&12&2&1&- 77 , 47&- 60 , 62&- 58 , NSP \\
\hline
20&1962 - 77&30 &16&4&- 4&- 47 , 70&- 50 , 72&- 42 , 76 \\
\hline
21&1973 - 84&32 &13&3&- 5&- 61 , NSP&- 61 , NSP&- 58 , NSP \\
\hline

\end{tabular}}

\end{table}

\renewcommand{\thetable}{B{2}}
\begin{table}[!ht]
\centering
\caption{Lists the phase difference in months between maximum activity at various latitude belts with respect to 35$^{\circ}$ belt for the southern hemisphere. ``NSP'' stands for when no significant peak is found in the cross-correlation curve.}
\label{tab:B2}
\vspace*{5mm}
\scalebox{0.7}{%
\begin{tabular}{ |c|c|c|c|c|c|c|c|c| }
 \hline
    \textbf{Cycle number} & \textbf{Period} &  \textbf{5$^{\circ}$} & \textbf{15$^{\circ}$} & \textbf{25$^{\circ}$} & \textbf{45$^{\circ}$} & \textbf{55$^{\circ}$} & \textbf{65$^{\circ}$} & \textbf{75$^{\circ}$}\\
\hline
                      \multicolumn{9}{|c|}{\textbf{EN area}} \\
\hline
15&1910 - 25&39&25&3&0&NSP,69&NSP, 67&NSP, 66\\
\hline
16&1921 - 36&30&10&3&- 3&NSP&NSP&NSP \\
\hline
17&1931 - 46&42&21&6&- 2&NSP&NSP&NSP \\
\hline
18&1941 - 56&32&18&5&- 2&- 26 , NSP&- 16 , NSP&- 30 , NSP \\
\hline
19&1952 - 67&34&20&4&- 3&- 43 , 80&- 42 , 76&- 33 , NSP \\
\hline
20&1962 - 77&35&21&2&- 2&- 49 , NSP&- 50 ,NSP&NSP \\
\hline
21&1973 - 84&28&15&3&- 1&NSP &- 34 , 63&- 34, 78 \\
\hline
                 \multicolumn{9}{|c|}{\textbf{AN area}} \\
 \hline
15&1910 - 25&38&22&5&-1&- 61 , 64&- 59 , 65&- 58, 67 \\
\hline
16&1921 - 36&35&12&3&0&NSP&NSP&NSP \\
\hline
17&1931 - 46&39 &21&5&- 2&NSP&NSP&NSP \\
\hline
18&1941 - 56&30&15&1&- 1&- 33 , NSP&- 33 ,  NSP&- 35 , NSP \\
\hline
19&1952 - 67&31&15&3&- 2&NSP&NSP&NSP \\
\hline
20&1962 - 77&33&15&0&- 2&- 53 , NSP&- 51 , NSP&- 37 , NSP \\
\hline
21&1973 - 84&26&13&4&0&NSP&NSP&NSP \\
\hline

\end{tabular}}

\end{table}

\section{Phase difference between  activity at different latitudes and sunspot number}

We have also analyzed the data for the individual solar cycles. Figure~\ref{fig:C1} shows the representative cross-correlation curves for the EN area for solar cycle number 19, considering the data for the period 1952--67.

\setcounter{figure}{0}
\renewcommand{\thefigure}{C{1}}

\begin{figure}[!ht]
\centering
\includegraphics[width=0.75\textwidth]{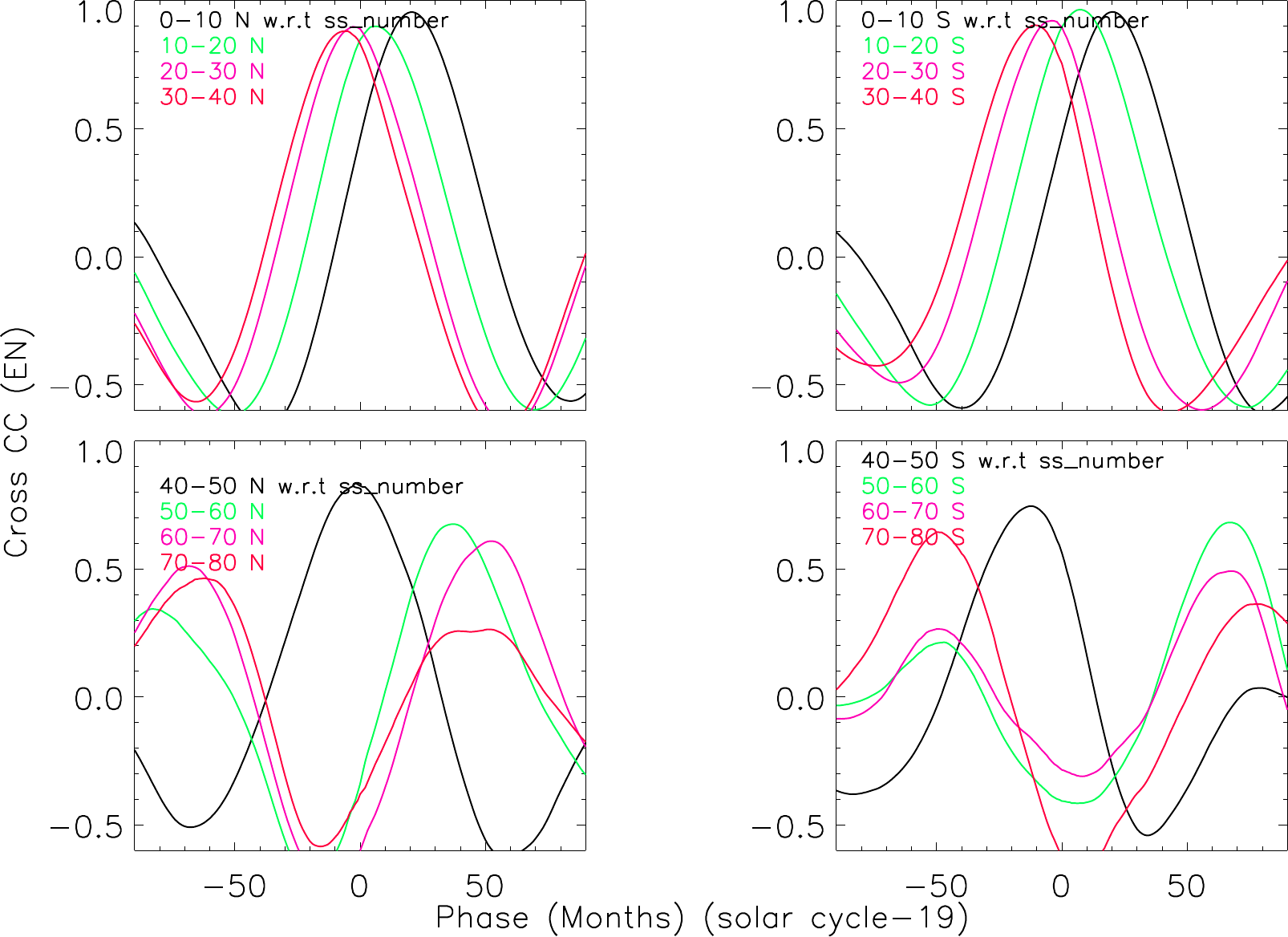}
\caption{Cross correlation curves of Ca-K EN area for 5$^{\circ}$, 15$^{\circ}$, 25$^{\circ}$, 35$^{\circ}$, and 45$^{\circ}$ latitude belts with respect to sunspot number (SS) over the visible disk considering the data period of 1952 -- 67 representing the analysis for solar cycle 19.}
\label{fig:C1}
\end{figure}

\setcounter{table}{0}
\renewcommand{\thetable}{C{1}}

\begin{table}[!ht]
\centering
\caption{Lists the phase difference in months between maximum activity at various latitude belts with respect to sunspot number for the northern hemisphere for each solar cycle. ``NSP'' stands for when no significant peak is found in the cross-correlation curve.}
\label{tab:C1}
\vspace*{5mm}
\scalebox{0.7}{%
\begin{tabular}{ |c|c|c|c|c|c|c|c|c|c| }
 \hline
   \textbf{Cycle number} & \textbf{Period} &  \textbf{5$^{\circ}$} & \textbf{15$^{\circ}$} & \textbf{25$^{\circ}$} & \textbf{35$^{\circ}$} & \textbf{45$^{\circ}$} & \textbf{55$^{\circ}$} & \textbf{65$^{\circ}$} & \textbf{75$^{\circ}$}\\
\hline
            \multicolumn{10}{|c|}{\textbf{EN area}} \\            
\hline
15&1910 - 25 & 14 & 0 &- 8 & -14 & -31,68 & NSP, 45 & NSP,48 & NSP,52 \\
\hline
16&1921 - 36&22&3&- 4&- 9&- 19 , NSP & - 62 , 81 & - 60 , 80 & - 55 , 83\\
\hline
17&1931 - 46&16 &3&- 2&- 6&- 7&- 28 ,  NSP&- 37 , NSP&NSP\\
\hline
18&1941 - 56&21&5&- 2&-8&- 12&- 41 , 65&- 45 , 62&- 43 , 64 \\
\hline
19&1952 - 67&22&6&- 2&- 7&- 2&- 81 , 37&- 67 , 54&- 60 , 49 \\
\hline
20&1962 - 77&11 &- 2&- 15&- 23&- 31&- 72 , 72&- 71 , 67&- 67 , 68 \\
\hline
21&1973 - 83&18 &2 &- 7&- 14&- 26&- 76 , 67&- 78 , 71&- 74 , 60 \\
\hline
  \multicolumn{10}{|c|}{\textbf{AN area}} \\
\hline
15&1910 - 25&13&0&- 6&- 13&NSP, 50&NSP, 46&NSP, 44&NSP, 51 \\
\hline
16&1921 - 36&22 &4&- 5&- 12&- 52 , NSP&- 63 , 82&- 54 , NSP&- 48 , 84 \\
\hline
17&1931 - 46&17&3&- 3&- 7& NSP&NSP&NSP&NSP \\
\hline
18&1941 - 56&21&4&- 4&- 8&NSP&- 41 , 64&- 45 , 64&- 44 , 68 \\
\hline
19&1952 - 67&21&5&- 4&- 8&0&- 77 , 40&- 68 , 56&NSP, 54 \\ 
\hline
20&1962 - 77&11&- 2&- 15&- 22&- 55&- 67 , 72&- 66 , 71&- 64 , 68 \\
\hline
21&1973 - 83&12&-3 &- 9&- 14&- 29&- 75 , 68&- 77 ,72&- 74 , 60 \\

\hline

\end{tabular}}

\end{table}

\renewcommand{\thetable}{C{2}}
\begin{table}[!ht]
\centering
\caption{Lists the phase difference in months between maximum activity at various latitude belts with respect to sunspot number for the southern hemisphere for each solar cycle. ``NSP'' stands for when no significant peak is found in the cross-correlation curve.}
\label{tab:C2}
\vspace*{5mm}
\scalebox{0.7}{%
\begin{tabular}{ |c|c|c|c|c|c|c|c|c|c| }
 \hline
  \textbf{Cycle number} & \textbf{Period} &  \textbf{5$^{\circ}$} & \textbf{15$^{\circ}$} & \textbf{25$^{\circ}$} & \textbf{35$^{\circ}$} & \textbf{45$^{\circ}$} & \textbf{55$^{\circ}$} & \textbf{65$^{\circ}$} & \textbf{75$^{\circ}$}\\
\hline
                        \multicolumn{10}{|c|}{\textbf{EN area}} \\
\hline
15&1910 - 25&23&8 &- 4&- 15&- 32 ,  NSP&- 43 , 72&- 60 , 50&- 69 , 65 \\
\hline
16&1921 - 36&19 &2&- 4&- 8&- 6&NSP&- 57 , 38&- 48 ,  63 \\
\hline
17&1931 - 46&19&5&- 6&- 14&- 18&- 25 , NSP&- 63 , NSP&- 60  , NSP \\
\hline
18&1941 - 56&18 &3&- 4&- 10&- 16&- 36 , 81&- 35 , 75&- 35 ,  73 \\
\hline
19&1952 - 67&20 &7&- 5&- 12&- 13&- 47 , 67&- 50 , 65&- 49 ,  78 \\
\hline
20&1962 - 77&21&8&- 6&- 14&- 22&- 76 , NSP&NSP&NSP ,  82 \\
\hline
21&1973 - 83&22&15&0&- 6&- 8&- 38 , NSP&- 42 , 48&- 40 , 72 \\
\hline
                 \multicolumn{10}{|c|}{\textbf{AN area}} \\
\hline
15&1910 - 25&22&8&- 4&- 13&- 29  , NSP&NSP , 52&- 63 , 50&- 64 , 63 \\
\hline
16&1921 - 36&21&4&- 4&- 8&NSP&NSP&- 62 , 40&- 54 , 65 \\
\hline
17&1931 - 46&19&3&- 6&- 13&- 18&- 26 , 75&- 61 , 67&- 66 , 65 \\
\hline
18&1941 - 56&18&2&- 4&- 10&- 20&- 40 , 75&- 48 , 72&NSP ,  72 \\
\hline
19&1952 - 67&20 &6&- 5&- 10&- 11&- 54 , 67&- 52 , 64&NSP \\
\hline
20&1962 - 77&22&8&-5&- 12&- 60&- 71 , NSP&- 73 , NSP&- 44  , NSP \\
\hline
21&1973 - 83&17&8&- 4&- 8&- 9&NSP&- 40 , 49&- 43 , 69 \\
\hline

\end{tabular}}
\end{table}

\end{document}